\documentclass[conference]{IEEEtran}
\IEEEoverridecommandlockouts

\PassOptionsToPackage{numbers,sort}{natbib}  
\usepackage{natbib}

\usepackage{hyperref}
\usepackage{url}

\usepackage{xcolor}
\definecolor{darkblue}{rgb}{0, 0, 0.5}
\hypersetup{colorlinks=true, citecolor=darkblue, linkcolor=darkblue, urlcolor=darkblue}

\usepackage[utf8]{inputenc} 
\usepackage[T1]{fontenc}    
\usepackage{url}            
\usepackage{booktabs}       
\usepackage{amsfonts}       
\usepackage{nicefrac}       
\usepackage{microtype}      
\usepackage{enumitem}
\setenumerate{topsep=0pt}
\usepackage{amsmath}
\usepackage{amssymb}
\usepackage{mathtools}
\usepackage{xfrac}
\usepackage{tabularx}
\usepackage{rotating}
\usepackage{tablefootnote}
\usepackage{amsthm}
\usepackage{bbm}
\usepackage{bm}
\usepackage{todonotes}
\usepackage{dblfloatfix}
\usepackage{float}
\usepackage{longtable}
\usepackage[most]{tcolorbox}

\usepackage[capitalize]{cleveref}
\usepackage{multirow}
\usepackage{makecell}
\usepackage{wrapfig}
\usepackage{tikz}
\usepackage{xspace}
\usepackage[]{caption}
\usepackage{adjustbox}

\usepackage{url}

\usepackage{breakurl}
\usepackage{xparse}
\usepackage{xcolor}
\usepackage{datasheet}

\newif\ifappendixonly
\appendixonlyfalse

\DeclareMathOperator*{\argmax}{arg\,max}

\newcommand{\itemlm}{{15pt}}
\newcommand{\benchmarkname}{\textsc{MarkMyWords}}

\newcommand{\mylink}[1]{\url{#1}}

\newcolumntype{z}[1]{>{\arraybackslash}m{#1}}   

\newcolumntype{R}[2]{%
  >{\adjustbox{angle=#1,lap=\width-(#2)}\bgroup}%
    c%
    <{\egroup}%
}

\newcommand{\rot}[1]{{\adjustbox{angle=60,lap=\width-1em}{#1}}}

\newcommand{\eat}[1]{}

\def\BibTeX{{\rm B\kern-.05em{\sc i\kern-.025em b}\kern-.08em
    T\kern-.1667em\lower.7ex\hbox{E}\kern-.125emX}}

\begin{document}

\title{\benchmarkname{}: Analyzing and Evaluating Language Model Watermarks }

\author{{\rm Julien Piet}\\ UC Berkeley
\and
{\rm Chawin Sitawarin}\\ UC Berkeley
\and
{\rm Vivian Fang}\\ UC Berkeley
\and
{\rm Norman Mu}\\ UC Berkeley
\and
{\rm David Wagner}\\UC Berkeley}
\maketitle

\begin{abstract}
  The capabilities of large language models have grown significantly in 
  recent years and so too have concerns about their misuse. 
  It is important to be able to distinguish machine-generated text 
  from human-authored content. 
  Prior works have proposed numerous schemes to watermark text, 
  which would benefit from a systematic evaluation framework.
  This work focuses on LLM output watermarking techniques---as 
  opposed to image or model watermarks---and proposes \benchmarkname{}, a 
  comprehensive benchmark for them under different natural language tasks.
  We focus on three main metrics: quality, size (i.e., the number of tokens needed to 
  detect a watermark), and tamper resistance (i.e., the ability to detect a watermark after
  perturbing marked text).
  Current watermarking techniques are nearly practical enough for real-world use:~\citet{kirchenbauer_watermark_2023}'s scheme 
  can watermark models like Llama 2 7B-chat or Mistral-7B-Instruct with no perceivable 
  loss in quality on natural language tasks, the watermark can be detected with fewer than 100 tokens, and their 
  scheme offers good tamper resistance to simple perturbations. However, they struggle to efficiently watermark code generations. 
  We publicly release our benchmark 
  (\mylink{https://github.com/wagner-group/MarkMyWords}).
\end{abstract}


\section{Introduction}
\label{sec:intro}

\footnotetext{Number of tokens needed to detect the watermark at a $p$-value of 0.02, as defined in~\cref{sec:metrics}.}

Recent advancements in large language models (LLMs) have been paralleled by escalating concerns over their misuse: automating social engineering attacks \citep{GovTech}, scaling propaganda operations \citep{goldstein2023generative}, and more \citep{gpt4system, salewski2024context, vasilatos2023howkgpt,wu2024fake,chen2023can}.
%
%
%
%

One approach to mitigate these risks is \emph{watermarking}, in which a subtle signal is embedded in all outputs from the model, so that others can detect LLM-generated text. To date, the most popular LLM watermarking techniques are \textit{symmetric-key} based, meaning a key is needed to encode the watermark into the LLM outputs and verify its presence.
%
%
%
%
\begin{figure}[th]
  \centering
  \includegraphics[width=\linewidth]{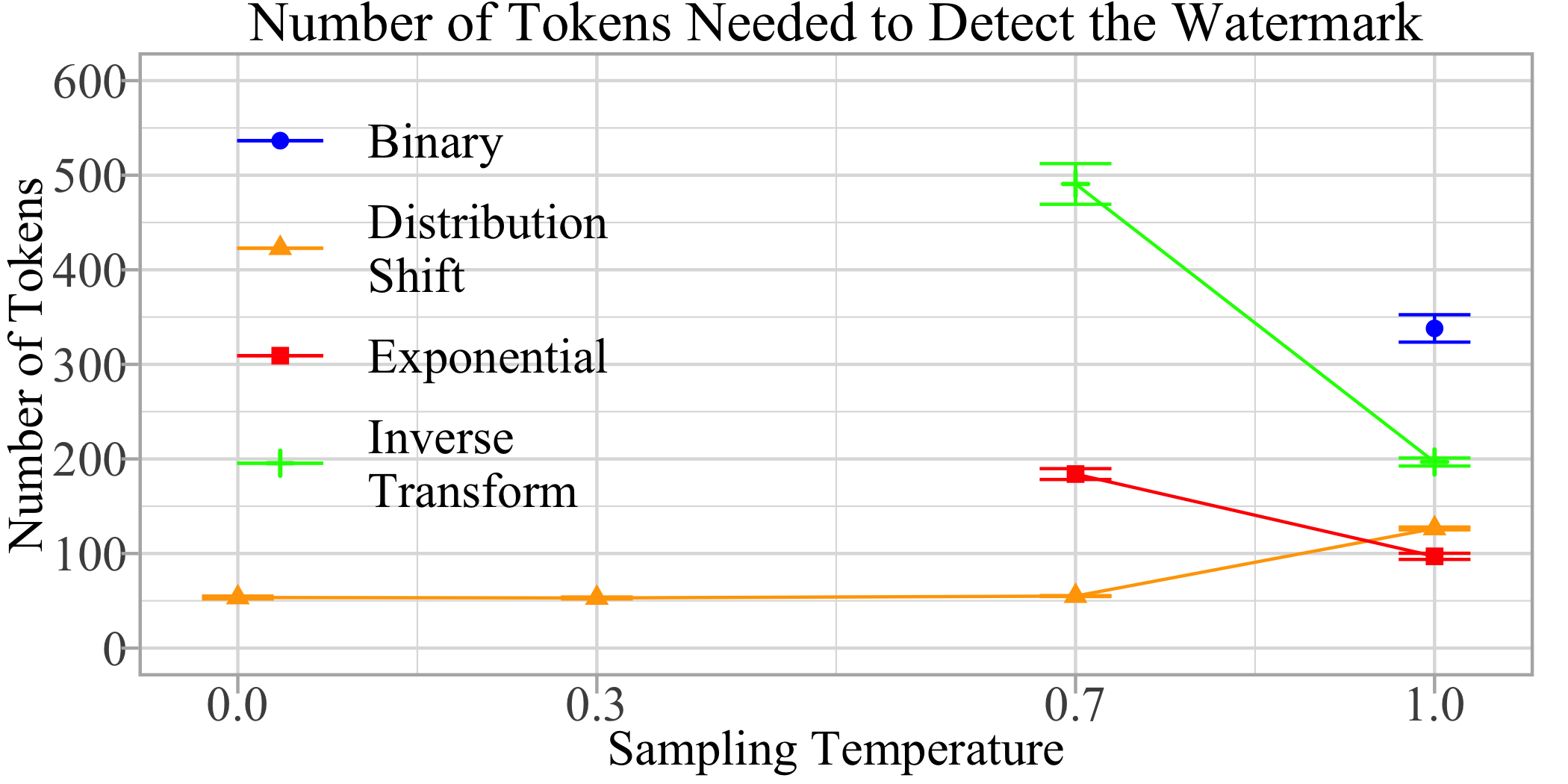}
  \caption[]{Watermark size\footnotemark~at near-optimal quality for four watermarking schemes (using Llama 2 7B-chat at various sampling temperatures). The distribution-shift scheme~\citep{kirchenbauer_watermark_2023} outperforms others at low temperatures, only needing a median of 60 tokens for the watermark to be detected.}
  \vspace{.1cm}
  \label{fig:aggregate}
\end{figure}
Multiple LLM output watermarking schemes have been proposed~\citep{aaronson_watermarking_2022,kirchenbauer_watermark_2023,christ_undetectable_2023,kuditipudi_robust_2023} and subsequently analyzed~\citep{chakraborty_possibilities_2023,sadasivan_can_2023,jiang_evading_2023,krishna_paraphrasing_2023}, but the feasibility of watermarking LLM outputs in practice remains unclear. Some researchers argue that watermarks can be practical \citep{krishna_paraphrasing_2023}, while others argue the opposite \citep{sadasivan_can_2023,jiang_evading_2023}.
There has yet to be consensus on evaluating different watermarking schemes or their readiness for practical deployment.

\begin{figure*}[t]
    \includegraphics[width=\linewidth]{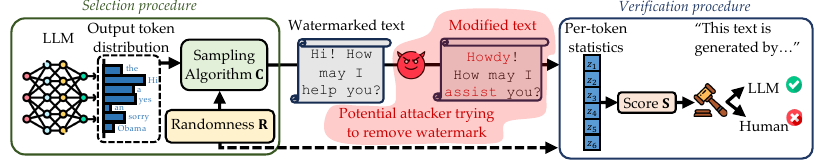}
    \caption{An overview of LLM-output watermarking.}\label{fig:diagram}
\end{figure*}

Our work tackles this challenge by providing a common ground where these algorithms can be empirically evaluated.
We propose \benchmarkname{}, an open-sourced benchmark to evaluate symmetric key watermarking schemes for under eleven tasks --- three realistic natural language tasks representing  possible misuses for selecting optimal watermarking parameters (book summarization, creative writing, and news article generation) and eight additional validation tasks to evaluate watermarks in various settings. 
We devise metrics to measure efficiency (number of tokens needed to detect a watermark), quality (whether the watermark degrades utility), and tamper resistance.
To measure tamper resistance, \benchmarkname{} tests whether a number of simple perturbations
can subvert the watermark without loss in quality.

We combine previous symmetric-key watermarking schemes into a unified framework, allowing practitioners to build custom schemes using building blocks from different prior work. We ran \benchmarkname{} on all practical parameter combinations and came to the following conclusions:
\begin{enumerate}[leftmargin=*,noitemsep]
    \item Watermarking schemes are nearing practicality for real-world use~(\cref{ssec:eval_aggregate}). The outputs of Llama 2~\cite{touvron_llama_2023} and Mistral~\cite{jiang2023mistral} can be watermarked with minimal quality loss for natural language tasks while detecting the watermark in under 100 tokens~(\cref{fig:aggregate}). However, optimal watermarks for natural language struggle to watermark code generation, incurring noticeable quality loss.
    \item Watermarks with optimal parameters are relatively robust to simple perturbations. Although more sophisticated attacks are capable of removing any watermark, GPT-3.5 paraphrasing only removes 50\% of the watermarks from the best scheme~(\cref{ssec:robustness-eval}). 
    \item We challenge the necessity of indistinguishability in natural language watermarking schemes, i.e., the distribution of watermarked outputs be provably indistinguishable from non-watermarked outputs~(\cref{ssec:discussion:indistinguishability}). Our results show \citet{kirchenbauer_watermark_2023} to be the easiest-to-detect watermark while maintaining quality, despite not being indistinguishable. 
\end{enumerate}

\noindent
The remainder of this paper is structured as follows. First, we provide some background on watermarking in~\cref{sec:background}. We provide details about our benchmark in~\cref{sec:experiments}, and introduce our metrics used to evaluate watermarks in~\cref{sec:metrics}. We present our findings in~\cref{sec:results}, and discuss implications and limitations of our work in~\cref{sec:discussion}. Finally, we provide an in-depth analysis of existing watermarking schemes and map out the design space in~\cref{sec:taxonomy}.

\section{Background}
\label{sec:background}

In this paper, we consider symmetric-key watermarking schemes that can be applied to existing pretrained language models.
They watermark the model's outputs, not the model's weights. 
These schemes are the most convenient and practical as they can be added to any generative language model
without requiring fine-tuning.
%
%
We provide here a high-level overview of the schemes.

\subsection{Related work}
\label{ssec:relwork}

Watermarking can refer to either watermarking 
{\it models}, or watermarking {\it model outputs}. Watermarking models~\citep{he2022cater, 
he2022protecting, cong2022sslguard, pmlr-v202-zhao23i} defends against model extraction attacks and is out of scope of our work; we focus on watermarking model outputs in order to detect AI-generated text.
Watermarking of textual data has been extensively 
studied~\citep{kamaruddin18,rizzo17,ahvanooey19,hopper2008stego}.
It can be viewed as a form of steganography~\citep{cox_information_2005,
majeed_review_2021, ziegler_neural_2019a} with a one-bit message. 
The message can be embedded post-generation (rule-based and neural-based watermarking), 
or during generation~\citep{tang2023science}. We consider the latter.
Watermarks could conceivably also be used 
to \emph{prove} that text was indeed machine-generated, for instance, 
to guarantee the provenance of a generation, similar to copyrights.
That setting is out of scope for this paper.

Another approach to detecting AI-generated text is to train a classifier on LLM outputs~\citep{openai_gpt2outputdataset_2021,openai_new_2023,gehrmann_gltr_2019a,mitchell_detectgpt_2023,bakhtin_real_2019,zellers_defending_2019a,ippolito_automatic_2020,
uchendu_authorship_2020,fagni_tweepfake_2021, NEURIPS2021_260c2432}.
This approach avoids the need to modify how text is generated by the LLM, but must be updated whenever the LLM changes.
For proprietary LLMs, an alternative is for the vendor to keep a copy of all generated 
outputs from their LLM and provide an API to look up whether a text was previously produced with their LLM~\citep{krishna_paraphrasing_2023}.



\smallskip\noindent\textbf{LLM benchmarks.}
New LLMs are accompanied by corresponding benchmarks designed 
to quantify their advancements over predecessors. While some benchmarks focus on 
assessing LLMs across a range of tasks---such 
as MMLU~\citep{hendrycks2021measuring}, BIG-Bench~\citep{yang2023toward}, and 
HELM~\citep{liang_holistic_2022}---others are tailored to evaluate specific 
capabilities like programming~\citep{austin2021program} 
or multi-turn conversation~\citep{zheng2023judging}. In contrast to these 
existing benchmarks, \benchmarkname~is the first to evaluate watermarks 
in LLM outputs across multiple dimensions: quality, efficacy, and tamper resistance. %
WaterBench~\citep{tu2023waterbench}, a concurrent benchmark, 
evaluates one class of watermarks \citep{kirchenbauer_watermark_2023}, 
and does not evaluate tamper resistance.
Furthermore, MarkMyWords better captures the ability of watermarking schemes 
to preserve quality. WaterBench selects hyperparameters that lead to detectable watermarks 
in generations of under 10 tokens, which in our experience is not a reasonable goal. 
The only way to achieve this is by significantly harming quality. We believe our approach---choosing parameters so that quality is preserved and then measuring how long outputs 
must be for the watermark to be detectable---is more realistic and informative. 
MarkLLM~\citep{pan2024markllm} is another concurrent watermarking framework focused on 
implementing and visualizing watermarking schemes. It provides a set of evaluation tools 
that tests each watermark using only one configuration of hyperparameters provided by the 
user. Instead, we evaluate hundreds of hyperparameter configurations and 
identify optimal trade-offs, as to fairly compare watermarking schemes 
and highlight the best performing one. 


\begin{table*}[t]
    \centering
    \small
    \begin{tabular}{llll}
    \toprule
        \multicolumn{2}{l}{\textbf{Randomness Source}} & \textbf{Description} & \textbf{Parameter}\\
    \midrule
        \multirow{2}{*}{\parbox{1.5cm}{\textbf{text-\\dependent}}} & sliding window & Keyed hash of sliding window of past tokens & \multirow{2}{*}{window size} \\
         & min hash & Minimum of keyed hash applied to each previous token in sliding window \\
     \midrule
         \textbf{fixed} & & Expands secret key to pseudorandom sequence & key length \\
    \bottomrule
    \end{tabular}
    \caption{Randomness sources.}
    \label{tab:randomness-sources}
\end{table*}

\subsection{Definitions}
\label{ssec:watermark-definition}

A (digital) watermark is a pattern embedded in a signal (image, text, audio, etc.) for 
identifying the source of the signal.
%
%
%
For generative language models, watermarks are useful
to detect machine-generated text, in contexts where using language models could be unethical, 
like phishing emails, fake news, or college essays.
These settings would benefit from a watermark that is not trivial for an unsophisticated adversary to remove. 
%
%

Watermarking algorithms consist of a marking procedure $\mathcal{W}$ 
and a verification procedure $\mathcal{V}$.
At generation time, each new token invokes the $\mathcal{W}$  to embed a watermark
into the generated output.
$\mathcal{W}$ has access to a secret key $k$, the previous tokens 
$T_0, \cdots, T_{n-1}$, and the language model's distribution 
$p(T_n \mid T_0,\dots,T_{n-1})$ on the next token, and selects 
a next token $T_n$.
$\mathcal{V}$ takes as input a secret key and a piece of text, and returns \texttt{True} if the text was generated using marking procedure $\mathcal{W}$.

\subsection{Evaluated watermarks}
\label{ssec:watermark-design}
We focus on symmetric-key watermarking algorithms.
To our knowledge, the only asymmetric key scheme was proposed 
by~\citet{fairoze2023publicly}. 
Existing watermarking schemes can be categorized by their source of randomness and sampling strategy for the next token.
Randomness sources can be combined with any sampling strategy.
There are 3 randomness sources used by prior work (\cref{tab:randomness-sources}), each using a keyed hash function. Text-dependent randomness sources rely on a fixed-size window of previous tokens. Fixed randomness depends only on the secret key. We identify 4 sampling strategies across prior work:
%

\vspace{1pt}\noindent\textbf{Exponential.} \citet{aaronson_watermarking_2022} mark text by selecting a token $T_n$ that 
    maximizes a score that depends on the probability $p(T_n|T_{0 \cdots n-1})$ and on a pseudorandom value $f_k(T_{n-H \cdots n-1})$ derived from a sliding window of $H$ prior tokens.
    %

\vspace{1pt}\noindent\textbf{Distribution-shift.}
\citet{kirchenbauer_watermark_2023} mark text by favoring tokens from a green list. 
    Green lists are derived from a pseudorandom value (computed either as $f_k(T_{n-1})$ or a min hash over a sliding window of $H$ prior tokens, $\min(f_k(T_{n-H}),\dots,f_k(T_{n-1}))$).
    %
    Tokens in the green list are favored by adding a small bias $\delta$ to their logits.\footnote{The green list size is denoted by $\gamma$~\citep{kirchenbauer_watermark_2023}. We set $\gamma=0.5$, and find other values of $\gamma$ degrade all metrics.}
    %

    \vspace{1pt}\noindent\textbf{Binary.} \citet{christ_undetectable_2023} convert the tokens to bit-strings.
    Each random bit is chosen based on a pseudorandom value (we use $f_k(T_{n-H \cdots n-1})$, derived from
    the LLM-induced distribution on bits).
    Finally, the bit-string is converted to a sequence of tokens.
    
\vspace{1pt}\noindent\textbf{Inverse transform.} \citet{kuditipudi_robust_2023} 
    computes the CDF $F(t) = \sum_{T_n=0}^t p(T_n|T_{0 \cdots n-1})$ of the LLM's output distribution (permuted according to a secret key $k$); then a fixed pseudorandom value $f_k(n)$ is used to sample from this distribution, via the inverse transform $F^{-1}(f_k(n))$.

These strategies are also used by other works:
\cite{takezawa2023necessary, zhao2023provable, lee2024wrote, liu2024unforgeablepubliclyverifiablewatermark, lu2024entropybasedtextwatermarkingdetection, lee2024wrotecodewatermarkingcode, liu2024semanticinvariantrobustwatermark, he2024watermarkssurvivetranslationcrosslingual, wu2024resilientaccessibledistributionpreservingwatermark} are based on distribution shift, 
and \cite{hu2024unbiased} uses inverse transforms. For the deterministic schemes above (all but distribution shift), we also experiment with adding more diversity in generated outputs by randomly skipping watermarking for some tokens with a given probability, the \emph{skip probability}.
We provide a full taxonomy unifying existing watermarking schemes in~\cref{sec:taxonomy}.

Verification algorithms rely on the likelihood of a score $\mathcal{S}$ under the 
hypothesis 
that the text was not watermarked. Watermarked text is flagged as AI-generated if the 
likelihood is below a $p$-value threshold. We also provide more details on verification algorithms in~\cref{app:ssec:score}.


\section{\benchmarkname{}}
\label{sec:experiments}




%
We now present \benchmarkname{}, our benchmark for evaluating watermarking schemes.
\benchmarkname{} focuses on natural language and relies on three text generation tasks, each comprised of about 100 examples.
They were chosen to generate long text, in order to ensure enough tokens to watermark most 
outputs and obtain a good size estimate (the number of tokens needed to detect a watermark, as defined in~\cref{ss:watermark-size}).
They represent scenarios in which LLM could be abused 
and thus in which watermarking would be useful. 
\begin{enumerate}[leftmargin=*,noitemsep,label=\textbf{(\arabic*)}]
    \item \textbf{Book reports.} Generate a report of a well-known book. (100 tasks)
    \item \textbf{Story generation.} Generate a short story, with a specific tone (e.g., funny, sad, suspenseful) and topic (e.g., ``three strangers that win a getaway vacation''). (96 tasks)
    \item \textbf{Fake news.} Generate a news story about two political figures meeting at an event. (100 tasks)
\end{enumerate}

%
%
%
%
%
%
Our benchmark generates a total of 296 outputs from the language model, 
with a maximum of 1024 tokens per generation.
We watermark the outputs of each task and measure quality and watermark size. 
We then perturb the generations to measure tamper resistance. We only attack the first third of each task to keep the benchmark runtime reasonable.
On an A5000 GPU, the benchmark completes 
within 40 minutes for one combination of a watermarking scheme and parameter setting.  Full task prompts are given in \cref{app:ssec:prompts}.

\smallskip \noindent{\bf Validation tasks.}
In addition to the three main tasks, we validate watermarks on eight additional tasks for a more holistic evaluation.

\begin{enumerate}[leftmargin=*,noitemsep,label=\textbf{(V\arabic*)},ref={V\arabic*}]
    \item \textbf{Domain-specific tasks.} Generate RFC\footnote{Requests For Comments (RFCs) are documents providing specifications for internet protocols.} summaries and legal research tasks. (50 tasks each) \label{validation-tasks:v1domain-specific}
    \item \textbf{Multilingual capabilities.} Generate book reports in French. (100 tasks) 
    \label{validation-tasks:v2-multilingual}
    \item \textbf{Low entropy tasks.} Paraphrase and translate book reports. (100 tasks each)
    \label{validation-tasks:v3-low-entropy}
    \item \textbf{Code generation.} Solve coding problems from the APPS dataset~\cite{hendrycks2021measuring}. (300 tasks)
    \label{validation-tasks:v4-codegen}
    \item \textbf{Short summarization.} Generate three sentence news highlights from the CNN/DailyMail dataset~\cite{see2017get, hermann2015teaching}. (100 tasks)
    \label{validation-tasks:v5-summarize}
    \item \textbf{Multiple choice.} Answer MMLU~\cite{hendryckstest2021} questions. (2000 tasks)
    \label{validation-tasks:v6-mmlu}
\end{enumerate}

\begin{table*}[th!]
    \centering
    \small
    \begin{tabular}{ll}
    \toprule
       \textbf{Perturbation Type}  & \textbf{Description} \\
    \midrule
        Swap & Randomly remove, add, and swap $p\%$ of the words in each sentence. \\
        Synonym & Replace $p\%$ of words in sentences with synonyms. \\
        Paraphrase & Use another LLM to paraphrase the output. \\
        Translation & Translate the output to another language and back to English. \\
        Contraction \& Expansion & Contract or expand verbs. \\
        Lowercase & Transform the output to all lowercase. \\
        Misspelling \& Typo & Add $p\%$ of typos or common misspellings. \\ 
    \bottomrule
    \end{tabular}
    \caption{Perturbations on watermarks included in \benchmarkname.}
    \label{tab:watermark-attacks}
\end{table*}

\subsection{Perturbations on watermarks}\label{sec:attacks}

Good watermarking schemes should easily detect the mark 
if an LLM's watermarked output is used directly.
Better schemes should also detect the mark even when the output 
is slightly modified.
Sufficiently sophisticated strategies can bypass any watermarking scheme~\citep{zhang2023watermarks}, but in many practical settings, this can be more effort than it's worth for the attacker: a cheating student trying to save time won't be inclined to carry out technically sophisticated attacks.
Therefore, we evaluate the watermarks against simple perturbations
aimed at removing the mark from AI-generated text. 
A perturbation of generated text $x$ is some text $x_{\mathrm{adv}}$ that is 
semantically similar to $x$, but can be syntactically different.
We summarize the attacks we use to provide a holistic 
evaluation of a watermark's tamper resistance in~\cref{tab:watermark-attacks} and provide more details below.

\smallskip\noindent\textbf{Swap attack.} 
%
One natural attack is to randomly remove, add, and swap some words in each sentence.
We scan generated text word by word, and with probability $p$, we either remove the word, duplicate it, or swap it with another randomly 
chosen word in the sentence.
%
%
Swap attacks are easy to implement for an attacker, and for small values of $p$ produce text that is still understandable.

\smallskip\noindent\textbf{Synonym attack.}
This attack replaces words in sentences with synonyms.
With probability $p$, we replace each word in the text by a semantically equivalent word.
This attack is more difficult to implement for an attacker.
We automate this attack using WordNet~\citep{miller_wordnet_1994} to zero-shot prompt GPT-3.5 to generate candidate synonyms.
%
%
In practice, this approach sometimes creates grammatically incorrect or unnatural sentences.
However, for a low probability $p$, the output text is still semantically 
close to the original.

\smallskip\noindent\textbf{Paraphrase attack.}
Perhaps the strongest attack in our toolkit, the paraphrase attack involves using another language model to rephrase the generated text.
This can be difficult and expensive to implement for an attacker, as they need access to a high-quality non-watermarked language model to do so, but the attack can completely change text without perturbing its meaning.
We implement two versions: (1) zero-shot prompting GPT-3.5 to 
paraphrase a generation, and (2) Dipper~\citep{krishna_paraphrasing_2023}, a fine-tuned model designed for paraphrasing.

\smallskip\noindent\textbf{Translation attack.}
This is similar to the paraphrase attack, except we use a translation model (\texttt{argos-translate}~\citep{finlay_argos_2021} based on OpenNMT~\citep{klein_opennmt_2017}) to translate text through a cycle of languages (e.g., English $\to$ French $\to$ English).
This attack does not alter the text as much as the paraphrase attack, but it is easy for an attacker to implement since they can use available services like Google Translate.
We use two languages, French and Russian, as variants of this attack.

\smallskip\noindent\textbf{HELM perturbations.}
HELM~\citep{liang_holistic_2022} implements a number of perturbations in its source code. 
They were originally designed to perturb model prompts. We use them to perturb model outputs. 
Among the list of perturbations they implement, we chose those that do not change the overall meaning of the text.
In particular, we use contractions \& expansions attacks, which contract verbs (\textit{e.g.} ``do not'' $\to$ ``don't'') or expand them, lower case attacks, which convert all words to lower case, misspelling attacks, which misspell each word with probability $p$, and typo attacks.

\subsection{Out-of-scope attacks}
\label{ssec:oos}
We evaluate the tamper resistance of schemes to simple perturbations.
We do not measure their robustness against stronger attackers and do not consider the following attack vectors:

\smallskip\noindent\textbf{Prompt modifications.} Some attacks modify the prompt to the model to avoid watermarking. 
For example, in the ``emoji attack'' the attacker instructs the model to insert an emoji between each word of the output, then replaces the emojis with spaces~\citep{kirchenbauer_watermark_2023}.
%
%
This attack defeats all watermarking schemes using text-dependent randomness. Prompt-modification strategies only work with models that can comprehend complex prompts, and can possibly be mitigated using advanced prompt filtering. 

\smallskip\noindent\textbf{Spoofing attacks.}
We do not consider attacks that spoof watermarks~\citep{jovanovic2024watermark,gu2024learnability} as the setting of proving provenance is out of scope for this paper.
Spoofing attacks can also enhance paraphrase attacks, but have a high one-time cost (e.g., 10,000s of generations) and can be mitigated by rotating the key $k$. 

\smallskip\noindent\textbf{Adaptive attacks.} We do not consider attacks that use the watermarking detection procedure $\mathcal{V}$ as an oracle. Mitigations include keeping the key $k$ secret, rate-limiting calls to $\mathcal{V}$, designing the verification API to release only ``watermarked'' or ``not'' (and not the score $\mathcal{S}$ or its likelihood), and detecting clusters of closely-related 
calls to the verification API.

\subsection{Implementation}
\label{ssec:implementation}

We implemented the \benchmarkname{}~benchmark in Python using the \texttt{transformers} library~\citep{wolf_transformers_2020} to implement models and watermarks.
%
Our code has been made public\footnote{\mylink{https://github.com/wagner-group/MarkMyWords}}.
It supports any language model available on HuggingFace and allows passing custom watermarking schemes for evaluating new solutions. 
%
We designed \benchmarkname{} with the goal of making future proposals of watermark schemes straightforward to evaluate.

\smallskip\noindent\textbf{Efficiency}
In order to speed up computation, we wrote custom implementations directly in CUDA of some of the watermarking schemes:
\begin{itemize}[leftmargin=\itemlm,nosep]
    \item \textbf{Hash function.} The exponential sampling scheme relies on computing the hash of many elements. Our CUDA implementation allows this to be done in parallel.
    \item \textbf{Edit distance.} The edit distance is computed with every possible key offset; our code implements this in parallel.
\end{itemize}

\smallskip\noindent\textbf{Reproducibility}
Our benchmark is packaged as a Python module, includes all necessary data, and can be installed easily by 
following the \texttt{README.md} file. Running the benchmark requires vLLM-compatible GPUs~\cite{kwon2023efficient}. The benchmark 
will produce deterministic results (quality, size and tamper-resistance) for a given randomness seed and 
watermarking secret key: only external components such as GPT paraphrasing are not fully deterministic, 
and are not part of the core benchmark. 
\section{Evaluation metrics}
\label{sec:metrics}

We propose three metrics for evaluating watermarking
schemes: (1) quality, (2) watermark size, and (3) tamper resistance.
We also propose an aggregate metric that summarizes the
performance of a watermarking scheme in a single number.

\subsection{Quality}

\benchmarkname~relies on a suite of tasks tailored for language models. Due to the scaling 
difficulties of human evaluation, we opt for automated ratings via LLM-as-a-
Judge~\citep{zheng2023judging,geval,chiang23rate,wang23rate,kocmi23rate}, despite potential biases 
such as preference for verbose answers and self-generated content~\citep{wang2023large}. Following 
\citet{zheng2023judging}, we mitigate these drawbacks by listing essential grading factors in the 
rating prompt (helpfulness, relevance, accuracy, depth, creativity, and level of detail) and using a different LLM for generation and rating. We provide the rating prompt in~\cref{app:ssec:prompts}. Our quality metric $Q$ is the average rating over all generations in the benchmark. 

We use Llama 3 (8B Instruct)~\citep{llama3} 
with greedy decoding as the judge LLM, for consistency and reproducibility.
\citet{zheng2023judging} shows that GPT-4 and GPT-3.5 produce ratings 
aligned with human preferences, and we found a high correlation ($R^2 > 0.9$
on our benchmark) between Llama 2 and GPT-3.5-Turbo or GPT-4-Turbo with our prompt.
We also found a high correlation between ratings from GPT-3.5-Turbo with our prompt versus GPT-3.5-Turbo with \citet{zheng2023judging}'s prompt.

We only use our quality metric to compare the relative quality of watermarked versus non-watermarked 
benchmarks. Its absolute value is not meaningful. We explored but ultimately dismissed 
model perplexity as a quality metric due to its preference for repetitive 
text~\citep{holtzman_curious_2020,welleck_neural_2020}. 

\benchmarkname~also uses the MAUVE score~\cite{pillutla2021mauve} as a secondary quality metric. 
MAUVE measures the distance between watermarked and non-watermarked distributions. 
Because of MAUVE's high variance on small datasets\footnote{MAUVE works best on distributions with $\geq$ 5000 samples.}, we only use it to validate the results obtained using 
our main quality metric.

\subsection{Watermark size}
\label{ss:watermark-size}

The longer the text, the easier it is to watermark and detect the watermark, as there are more degrees of freedom to inject a mark.
Therefore, a critical metric is: how long must the generated text be, so that we are likely to be able to detect the watermark in it?

The verification algorithm can make two types of errors: false positives (when an unwatermarked text is detected) and false negatives (when a watermarked text is not detected).
%
All schemes we consider rely on one-tailed statistical tests, so we can precisely control the false positive rate (by setting the $p$-value threshold).
We define the watermark size, $E$, to be \textbf{the number of tokens needed to detect the watermark, at a 2\% false positive rate}.
We measure it by finding the shortest prefix detected as marked on each output ($+\infty$ if no prefix is detected) and then computing the median of the lengths.
Smaller values of $E$ indicate better, more efficient watermarking schemes.



\subsection{Tamper resistance}
\label{ssec:tamper-resistance}

We assess the robustness of watermarking schemes against 8 basic tampering attacks outlined in \cref{sec:attacks}. Each attack's impact is quantified by measuring both the quality retention ($Q_A$, indicating the extent to which an attack degrades the output quality) and the detection rate ($W_A$, reflecting the percentage of generations still watermarked after perturbation\footnote{We use this instead of the watermark size because many attacks remove the watermark from over 50\% of generated texts, in which case all these attacks would have a size of $+\infty$.}).

We define $Q_A:=\max(0,\min(Q^*_A/Q,1))$, the clipped ratio of the mean quality of attacked outputs $Q^*_A$ to that of the baseline.
Experimentally, all attacks except the ``contraction'' attack substantially modify the output and reduce quality.
Contraction attacks may leave outputs unchanged, sometimes exhibiting quality up to 1\% higher than the baseline due to variance.

\begin{figure}
\centering
\includegraphics[width=\linewidth]{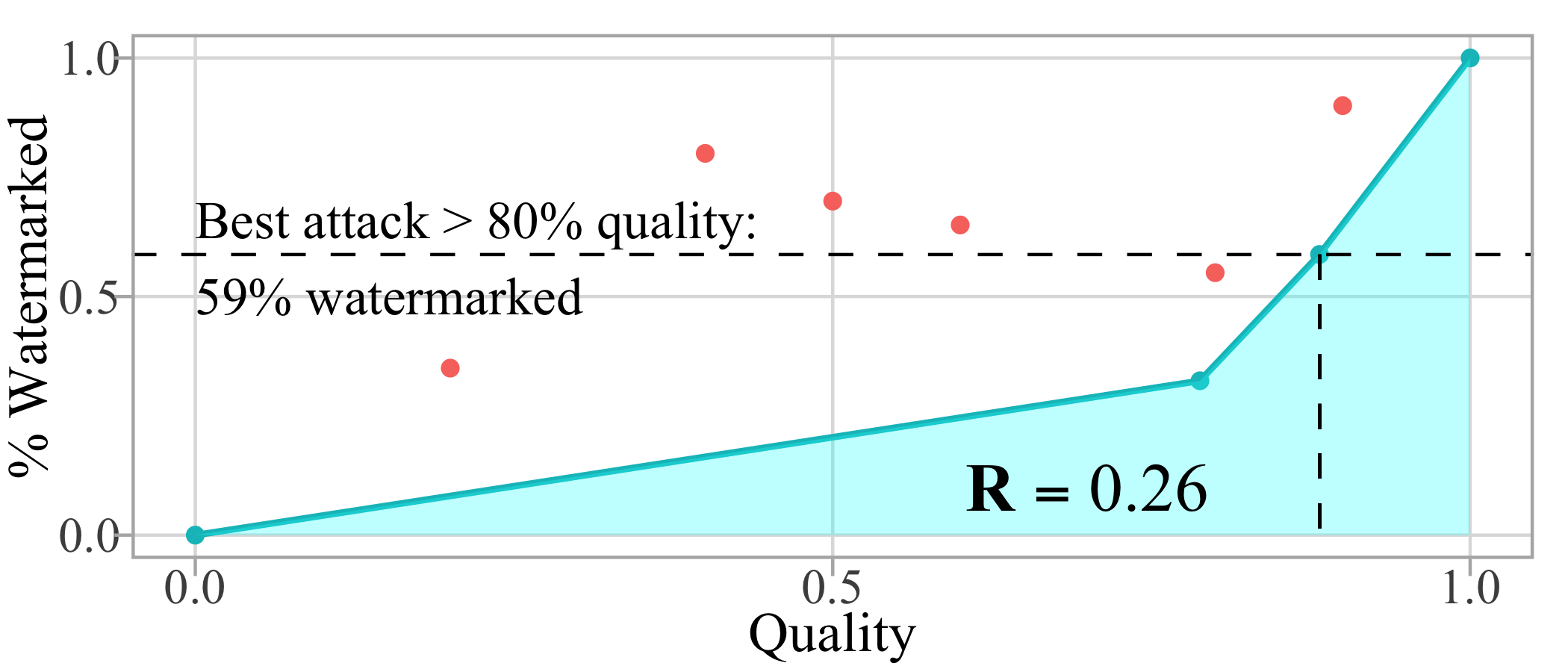}
\vspace{-0.5cm}
\caption{Example tamper resistance plot. Blue points are attacks on the \emph{convex hull's frontier}. The blue area represents the AUC and the dashed lines the best attack preserving 80\% quality. }
\label{robustness-example-fig}
\end{figure}

We can visualize attacks by plotting their quality retention vs. detection efficiency (\cref{robustness-example-fig}).
The closer a point is to $(1,0)$, the more successful the attack.
$(1,1)$ corresponds to no attack, and $(0,0)$ is a trivial attack that returns an empty string.
We define the tamper resistance of a watermark as the area under the curve (AUC) derived from the convex hull, representing the attacks with the best possible trade-off between retained quality and effectiveness. Any point on the hull's frontier is attainable (on average) by sampling between the two closest attacks on the hull. A value of 0 means a tamperable scheme, and a value of 0.5 is the highest achievable tamper-resistance. We define $R$ as twice the AUC, so $R$ is normalized to be between 0 and 1. This metric has a similar intuition to the area under the curve measured in ROC curves in binary classification~\citep{fawcett_introduction_2006}.

We exclude paraphrasing attacks when computing $R$, because they rely on large or closed-sourced language models, which can be difficult to obtain, require expensive resources to run, or can be watermarked themselves. We do include translations, as they rely on smaller models and are freely accessible online. We evaluate paraphrasing attacks in~\cref{ssec:robustness-eval}.

\begin{figure}[h]
    \includegraphics[width=\linewidth]{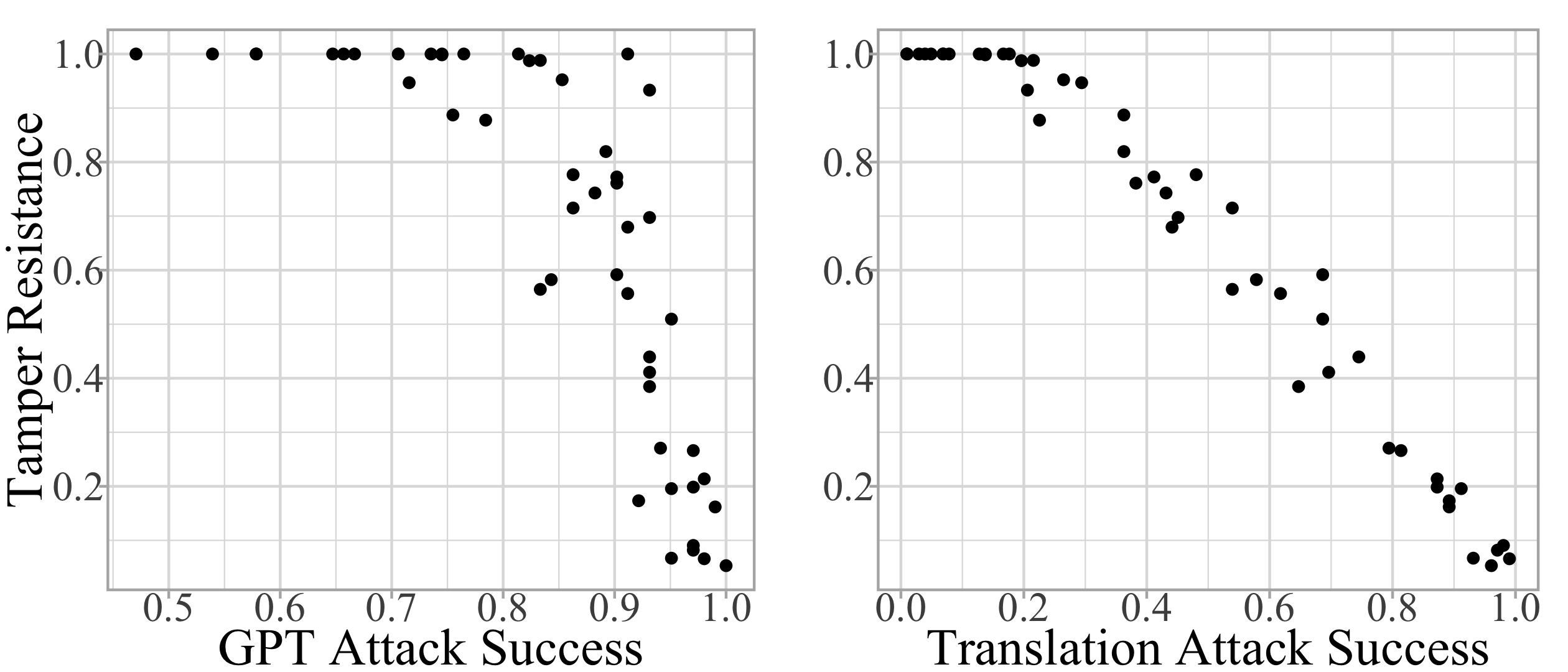}
    \centering
    \caption{Correlation between robustness metric and attack success. On the right, for the Russian translation attack. 
    On the left, for the GPT paraphrasing attack. Each point is a unique watermark parameter setting.}
    \label{fig:robustness-to-attacks}
\end{figure}

We found $R$ to be correlated to the success rate of different attacks (see ~\cref{fig:robustness-to-attacks} in \cref{app:ssec:additional_figures}). 
We found that $R=1$ corresponds to $<20\%$ success rate for the Russian translation attack and $<70\%$ for the GPT paraphrase attack. 

\subsection{Aggregate metric}\label{ssec:aggregate}

Different settings of a watermark's parameters~(\cref{ssec:param_tuning}) result in different metrics.
In order to compare two watermarking schemes \emph{overall}, we propose an aggregate metric.
We first set the watermark's parameters to be optimal with respect to watermark size on a training set while achieving a target quality and tamper resistance. The aggregate metric is this watermark's size when run with different random seeds and secret keys, to avoid selection bias. 

\cref{fig:aggregate} reports this aggregate metric for all four schemes,
with a target quality degradation of $\leq 1\%$ and target tamper resistance of $>0.2$.
%
\cref{robustness-per-scheme} and \cref{fig:detailed_llama_values} present results 
for other thresholds (optimizing for size vs. tamper resistance, $\leq 1\%$ vs. $\leq 10\%$ quality degradation, $>0.2$ vs. $>0$ tamper resistance). 

\section{Results}
\label{sec:results}

\begin{figure}[t!]
    \centering
    \includegraphics[width=\linewidth]{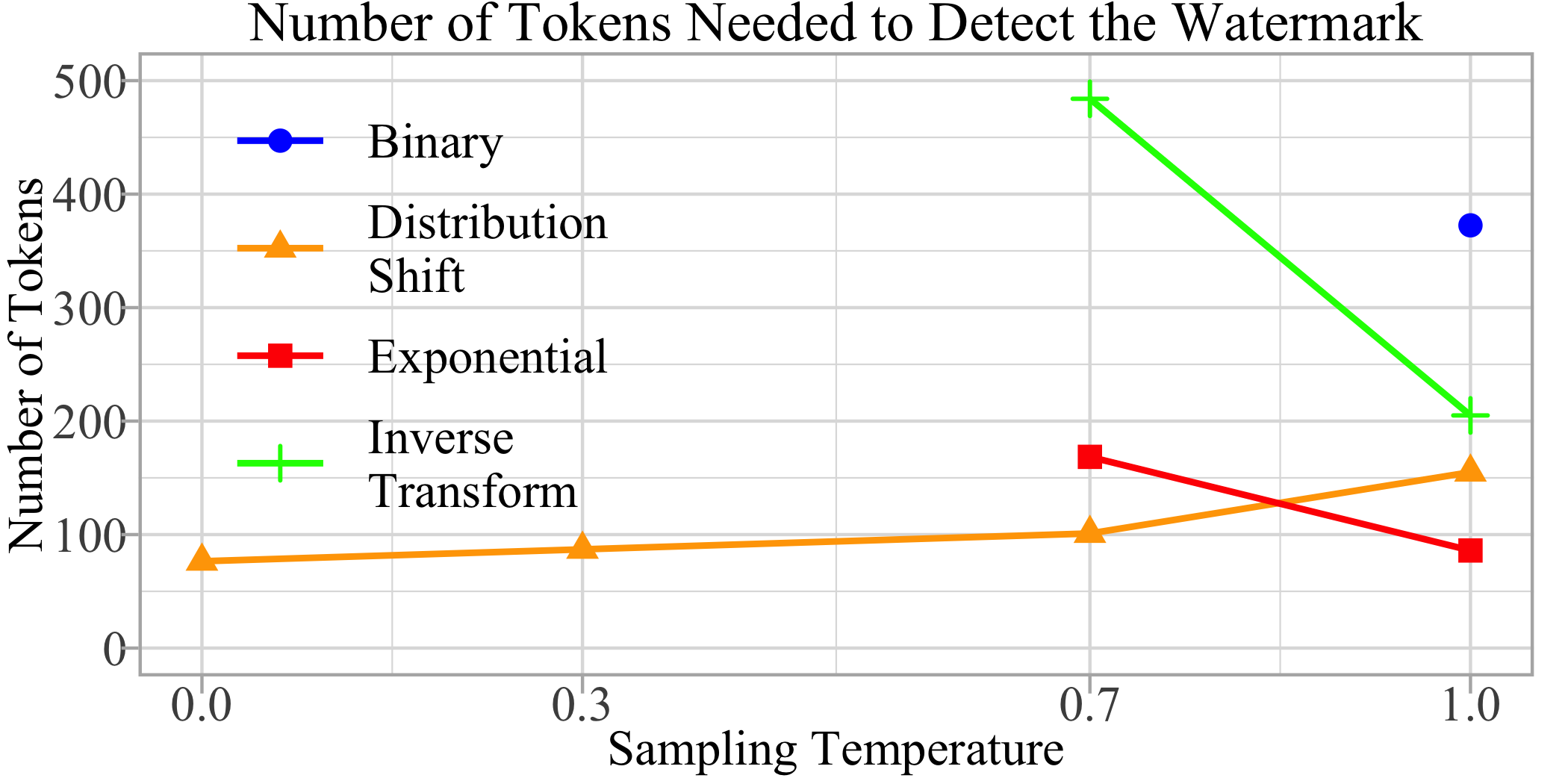}
    \begin{minipage}{0.48\textwidth}
        \caption{Watermark size at near-optimal MAUVE quality for each watermarking schemes taken from the literature, using Llama 2 7B-chat at various sampling temperatures.}
        \label{fig:mauve_results}
    \end{minipage}
    \vspace{0.5cm}
\end{figure}

We evaluated more than 1,200 unique combinations of watermark schemes and 
parameter settings. This took $\sim$1 week to run on 4 A5000 GPUs. Our results suggest 
using a distribution-shift sampling strategy with text-dependent randomness 
(\cref{ssec:eval_aggregate}).  We give some parameter recommendations in~\cref{ssec:param_tuning}, 
and evaluate tamper resistance in~\cref{ssec:robustness-eval}.

\subsection{The ``best'' watermark}\label{ssec:eval_aggregate}
In \cref{fig:aggregate}, we show the minimal watermark size (median number of tokens needed to detect a watermark, at 2\% false positive rate) under various temperatures, for a quality degradation of at most 1\% and achieving tamper resistance of at least $R > 0.2$.
The distribution shift scheme performs the best in the 0.0--0.7 temperature range, 
which is arguably the most common range of temperature values used in practice.\footnote{For example, GPT-4's technical report uses a temperature of 0.6~\citep{achiam2023gpt-techrep}.}
At temperature 1, exponential sampling is slightly superior to distribution shift.
The relative ranking of watermarking schemes is consistent across 
different choices of quality, tamper resistance 
threshold, and quality metric (MAUVE vs. Llama 3 ratings), as shown in~
\cref{fig:mauve_results}, an analogue of ~\cref{fig:aggregate} using MAUVE as a quality metric. We define near-optimal quality to be a MAUVE similarity within 2.5\% of the baseline, as stricter bounds are not meaningful since the empirical standard deviation of MAUVE scores on our dataset is 1.3\%. 

\begin{figure}[t]
    \centering
    \includegraphics[width=\linewidth]{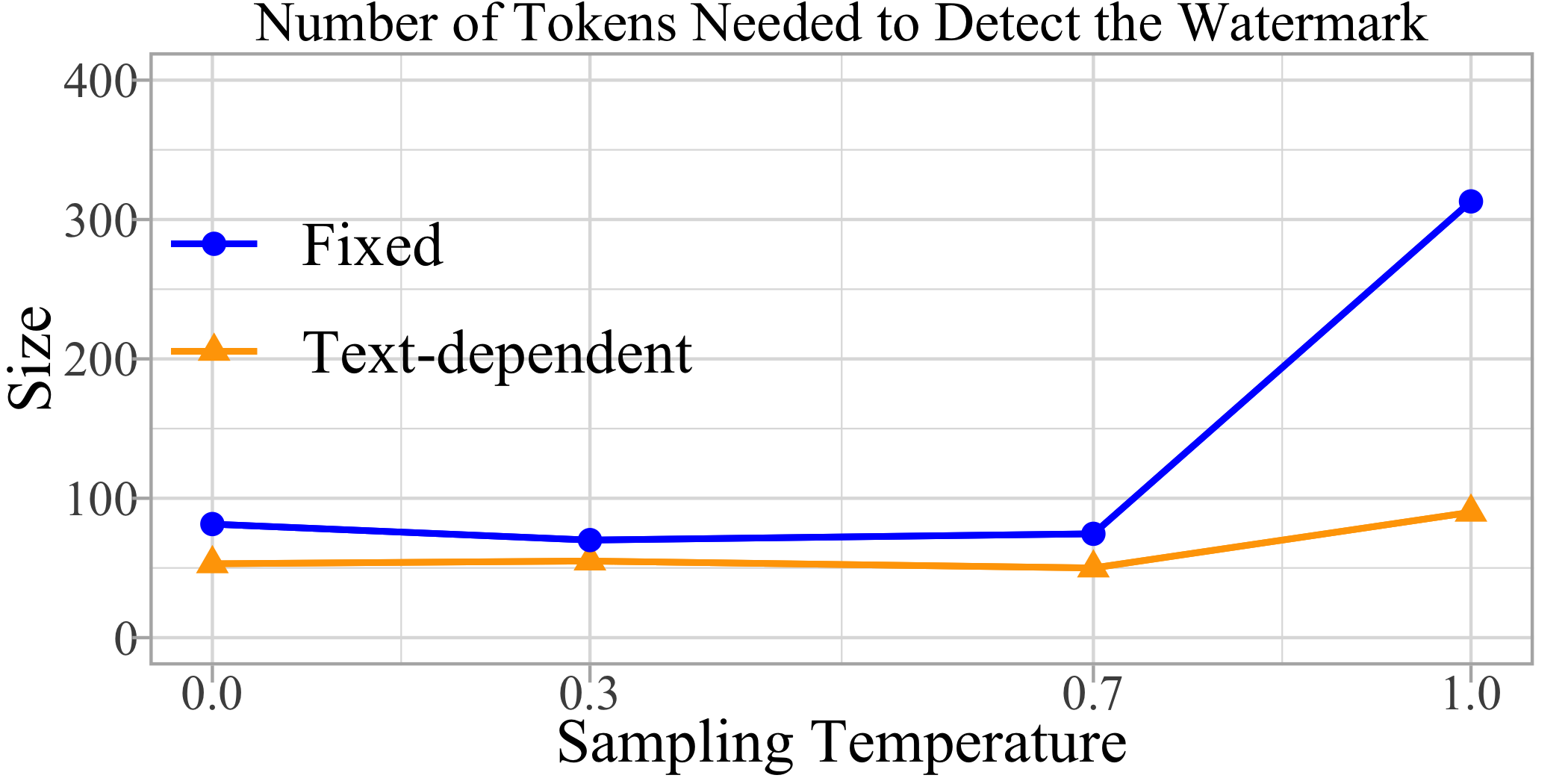}
     \caption{Watermark size at near-optimal quality for text-dependent versus fixed randomness.  The values correspond to the minimal size of schemes with near-optimal quality. Text-dependent randomness is more efficient at all temperatures.}
    \label{fig:text-dependent_fixed_randomness}
\end{figure}

\smallskip\noindent{\bf Ready to deploy?} 
The distribution shift scheme is able to detect watermarks with
$\sim$50--60 tokens (roughly 40 words), at temperatures in the
range 0.0--0.7.
This suggests that the distribution shift watermarking scheme
is practical enough to be deployed today {\bf for natural language generations}.

\begin{figure}[t]
    \centering
    \includegraphics[width=\linewidth]{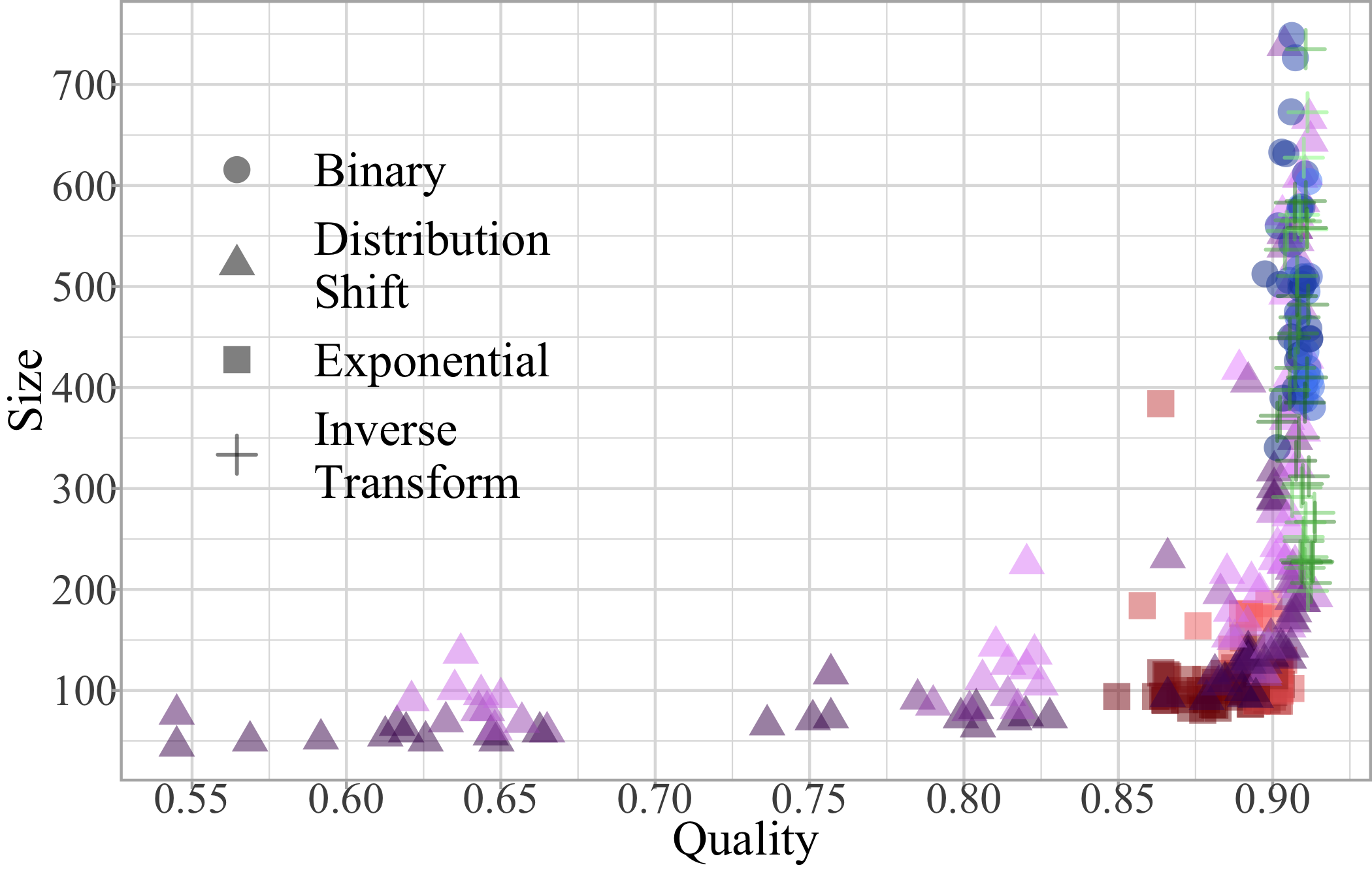}
    \caption{Size versus quality of all parameter settings at T=1. The darker the color, the more tamper-resistant. Distribution shift is the most tunable scheme.}
    \label{fig:all_params}   
\end{figure}

\begin{figure*}
    \centering
    \begin{minipage}{.49\textwidth}
        \centering
        \includegraphics[width=\linewidth]{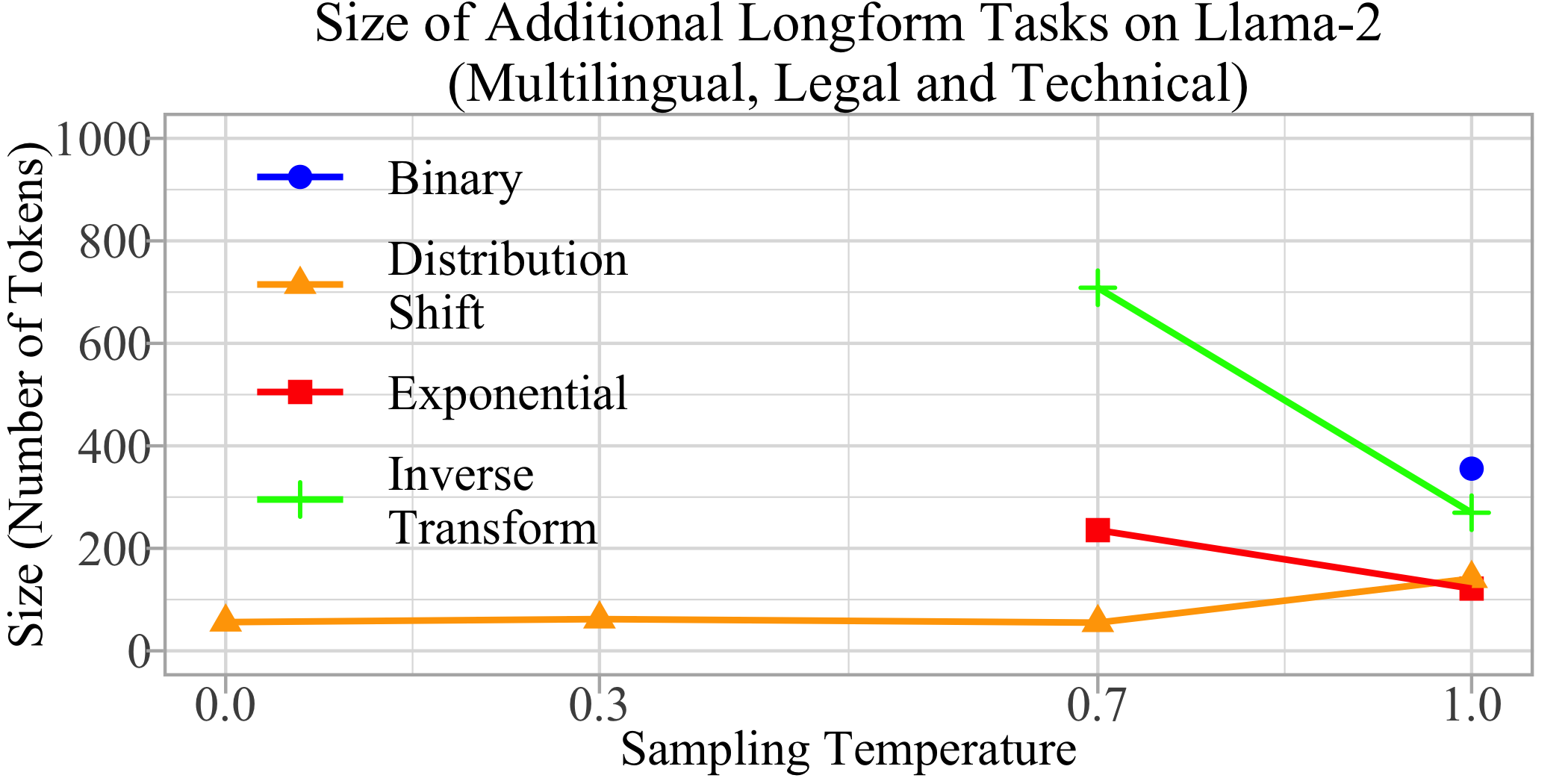}
    \end{minipage}
    \hfill
    \begin{minipage}{.49\textwidth}
        \includegraphics[width=\linewidth]{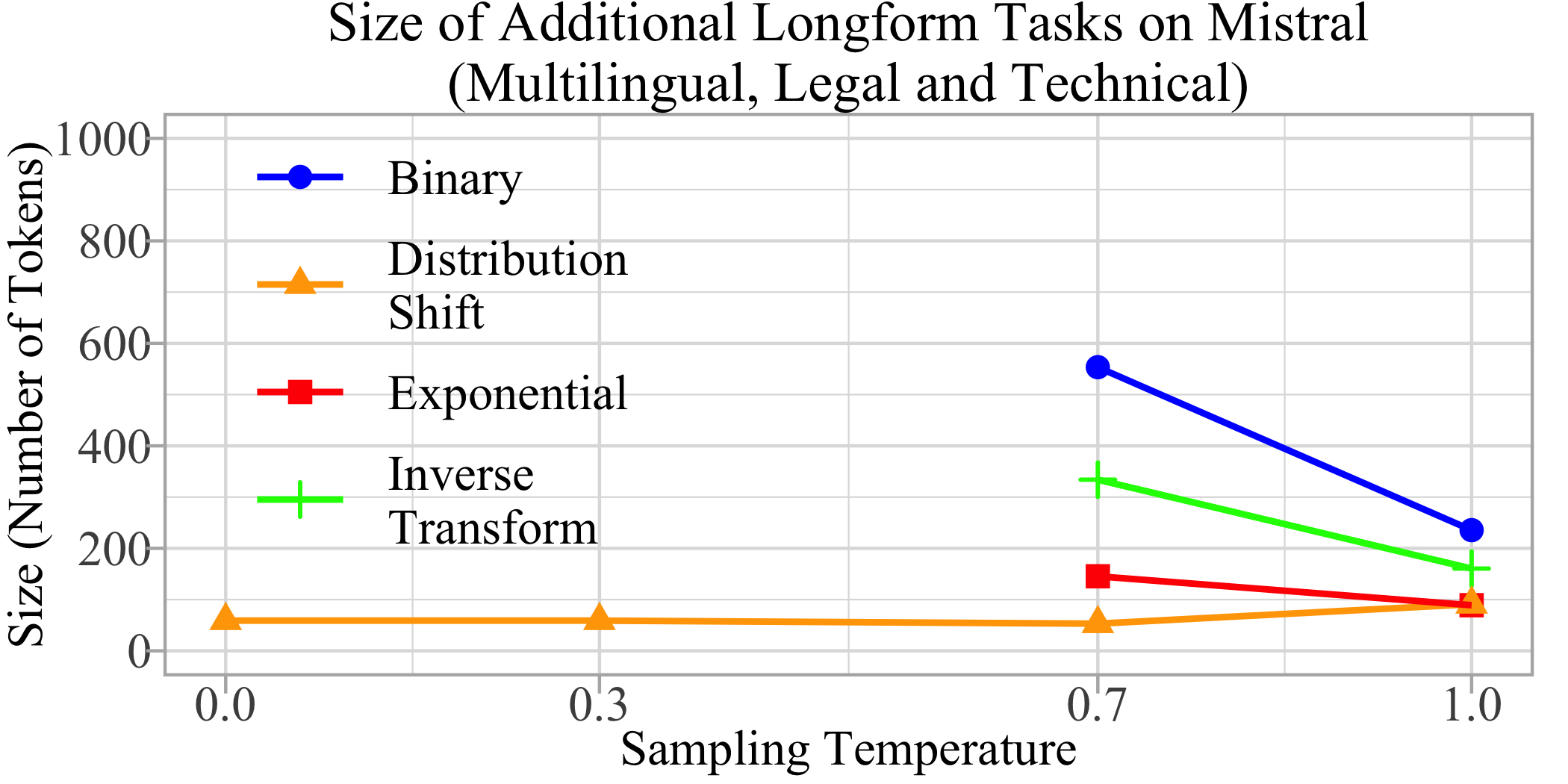}
        \centering
    \end{minipage}
    \caption{Size of watermarks from~\cref{fig:aggregate} on validation tasks V1 and V2 both Llama-2-7B-chat and Mistral-7B-Instruct.}
    \label{fig:add_tasks_size}
\end{figure*}

\begin{figure*}
    \centering
    \begin{minipage}{.49\textwidth}
        \centering
        \includegraphics[width=\linewidth]{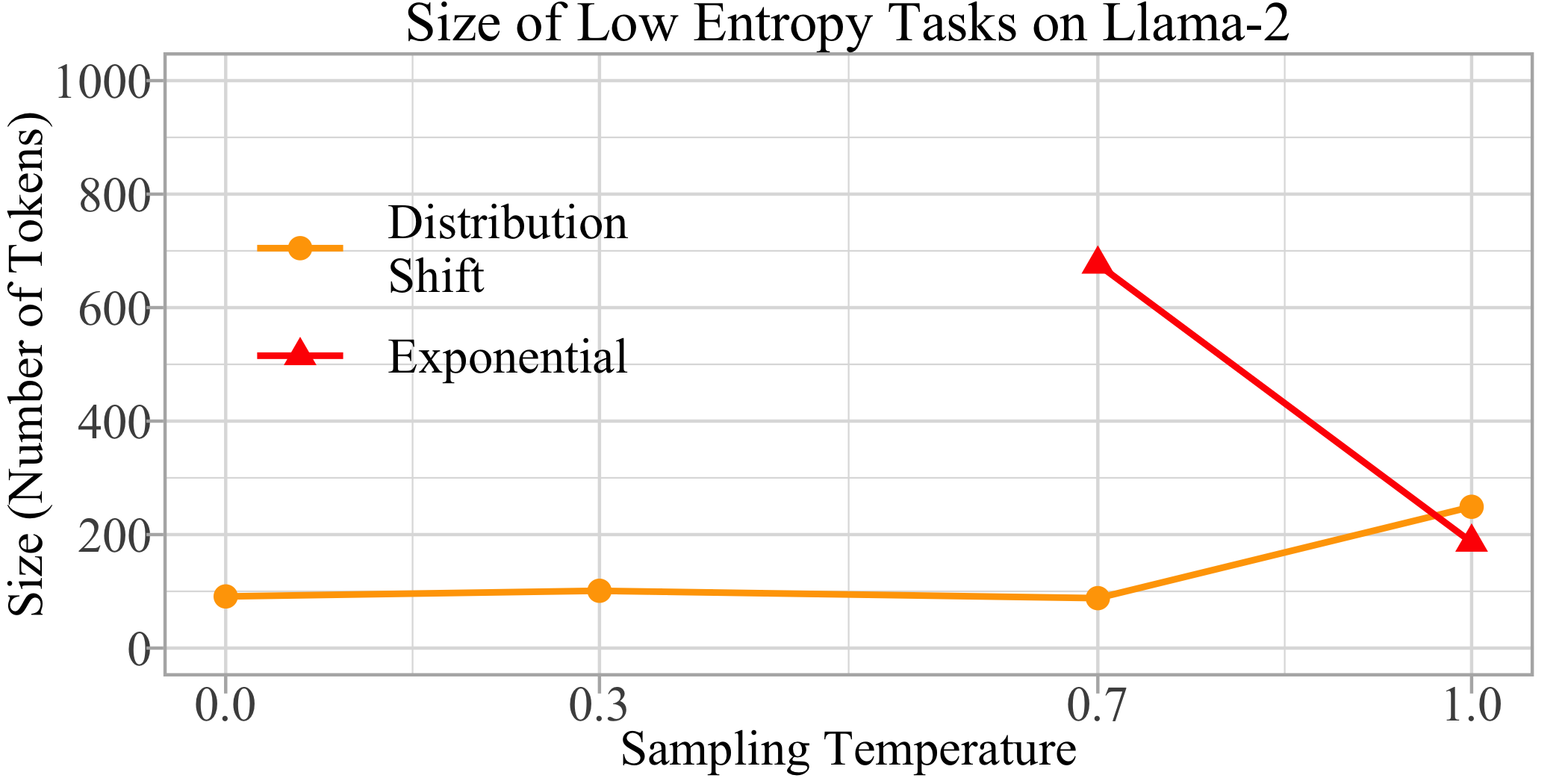}
        \caption{Watermark size using watermarks from~\cref{fig:aggregate}.}
        \label{fig:lowentropy-size}
    \end{minipage}
    \hfill
    \begin{minipage}{.49\textwidth}
        \includegraphics[width=\linewidth]{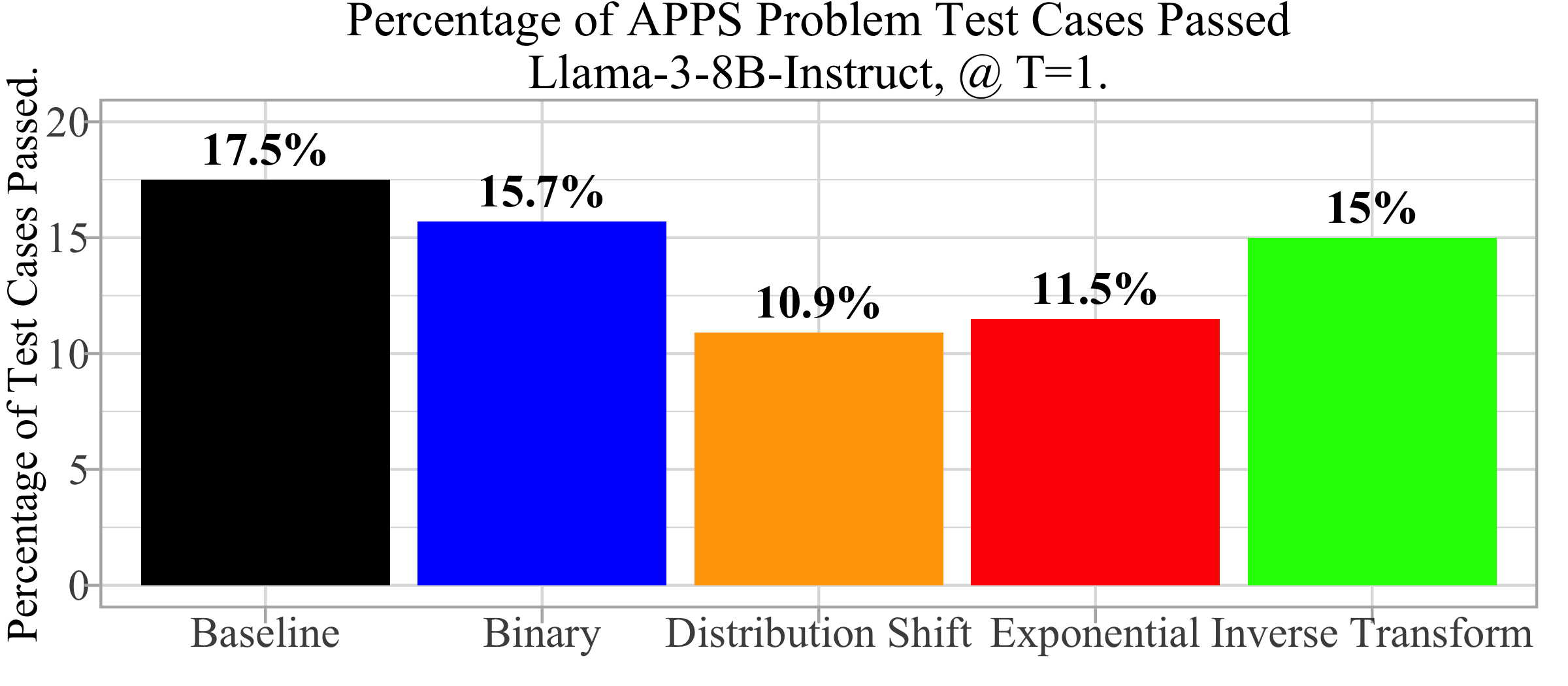}
        \centering
        \caption{Figure R5: Percentage of APPS test case problems passed, using Llama-3-8B-Instruct at T=1 with the optimal schemes for Llama-2.}
        \label{fig:code-quality}
    \end{minipage}
\end{figure*}

\smallskip\noindent\textbf{Validation tasks.} Performance on domain-specific tasks as well as multilingual tasks (\ref{validation-tasks:v1domain-specific} and \ref{validation-tasks:v2-multilingual}) is nearly identical to the three main tasks, as shown in~\cref{fig:add_tasks_size}, while preserving quality. The same watermarks achieve good quality on the paraphrasing and translation tasks (\ref{validation-tasks:v3-low-entropy}). Watermark size increases by a factor of 2 to 3 due to the lower entropy. Distribution shift watermarks can still be detected in under 100 tokens for $T \leq 0.7$, and in 250 tokens at $T=1$. More details can be found in~\cref{fig:lowentropy-size}. Short summaries (\ref{validation-tasks:v5-summarize}) and Q\&A responses (\ref{validation-tasks:v6-mmlu}) are too short for watermarks to be detected. We focused on evaluating the quality of the generations. Watermarking preserved the quality of the short summaries, but decreased quality of both Q\&A responses and code generations (\ref{validation-tasks:v4-codegen}). 

\begin{figure*}
    \centering
    \begin{minipage}{.49\textwidth}
        \centering
        \includegraphics[width=\linewidth]{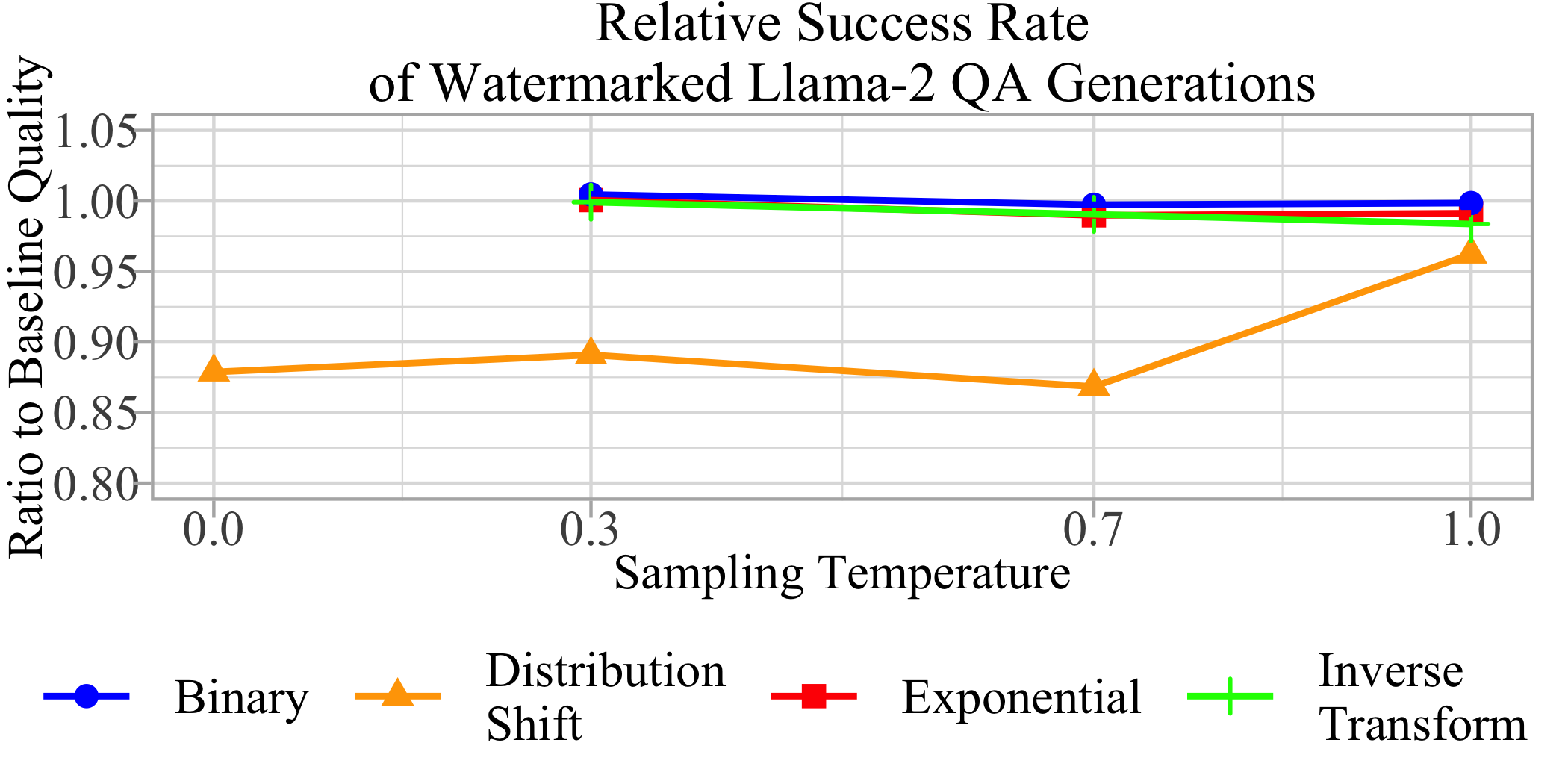}
    \end{minipage}
    \hfill
    \begin{minipage}{.49\textwidth}
        \includegraphics[width=\linewidth]{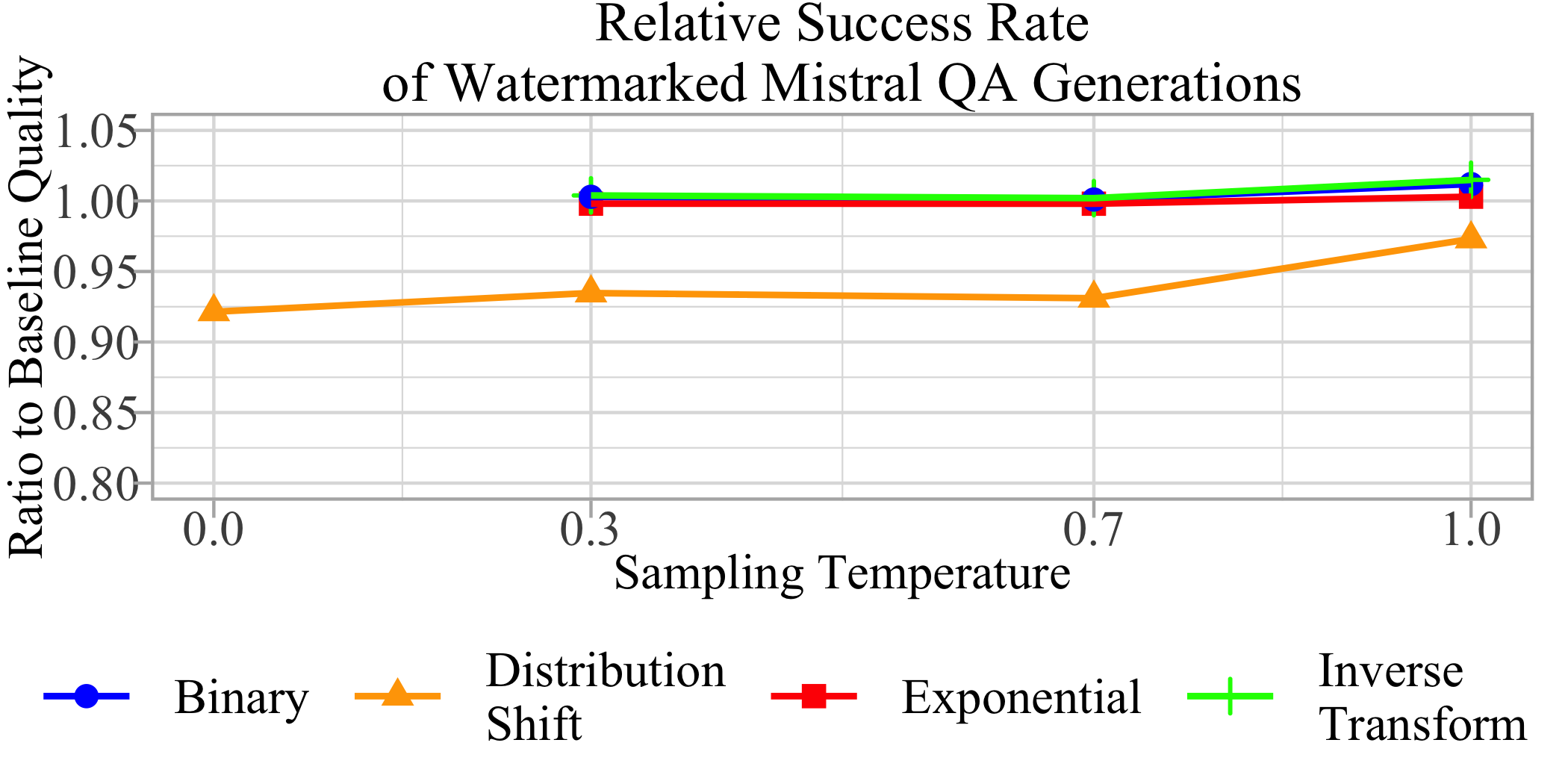}
        \centering
    \end{minipage}
    \caption{Correctness of watermarked QA answers (V6) relative to the correctness of the non-watermarked baselines, for watermarks from~\cref{fig:aggregate}.}
    \label{fig:qa_tasks_quality}
\end{figure*}

\smallskip\noindent\textbf{Structured outputs.} Watermarks struggle to preserve the quality of structured outputs, like code generation or multiple choice answers. The optimal watermarks from~\cref{fig:aggregate} decrease the quality of responses to coding tasks (\ref{validation-tasks:v4-codegen}). At a temperature of $T=1$, using Llama-3-8B-Instruct~\cite{llama3}\footnote{Llama2 and Mistral did not perform well on this task.}, non-watermarked generations pass 17.5\% of the APPS test suite, while distribution shift scheme generations pass 11\%, and exponential scheme generations pass 11.5\% (\cref{fig:code-quality}). Code generations have lower entropy and are shorter than the natural language generations in our benchmark. The distribution-shift scheme only identifies 40\% of generations while the exponential scheme identifies 58\%, with a watermark size of 200. We observe a similar phenomenon on MMLU answers (\ref{validation-tasks:v6-mmlu}), where the distribution shift parameters from~\cref{fig:aggregate} make 5-10\% more mistakes than the baseline for Llama2 and 2-7\% for Mistral. However, other schemes were able to perform MMLU tasks with no decrease in accuracy. We plot the correctness of watermarked Q\&A answers in~\cref{app:ssec:additional_figures}  (\cref{fig:qa_tasks_quality}).

\smallskip\noindent{\bf Parameter settings.} 
\cref{fig:aggregate} uses the following parameters: for distribution-shift, 
text-dependent randomness with min hash of size 3 ($T=0.7$), sliding window of size 1 ($T\neq 0.7$), $\delta=5$ ($T\leq0.7$), and $\delta=2.5$ ($T=1$);
for the exponential scheme, sliding window size $3$, and skip probability $0.05$. 
These achieve our goals of $R>0.2$ tamper resistance and $\leq 1\%$ quality degradation.

\smallskip\noindent\textbf{Randomness source.} Text-dependent randomness is more efficient than fixed randomness. Fixed randomness requires between 20 to 200 additional tokens for detection, depending on the sampling temperature, as seen in~\cref{fig:text-dependent_fixed_randomness}, in which we report the smallest size result across all four schemes for fixed randomness (in blue) versus text-dependent randomness (in orange). For text-dependent randomness, we report the smallest size between sliding window and min hash.\footnote{We exclude text-dependent randomness with a window size of 0 or fixed randomness with a key length of 1 from the analysis, since these corner cases are identical and correspond to always using the same random value.}

\begin{figure*}[h]
    \begin{minipage}{0.6\textwidth}
        \centering
        \includegraphics[width=\linewidth]{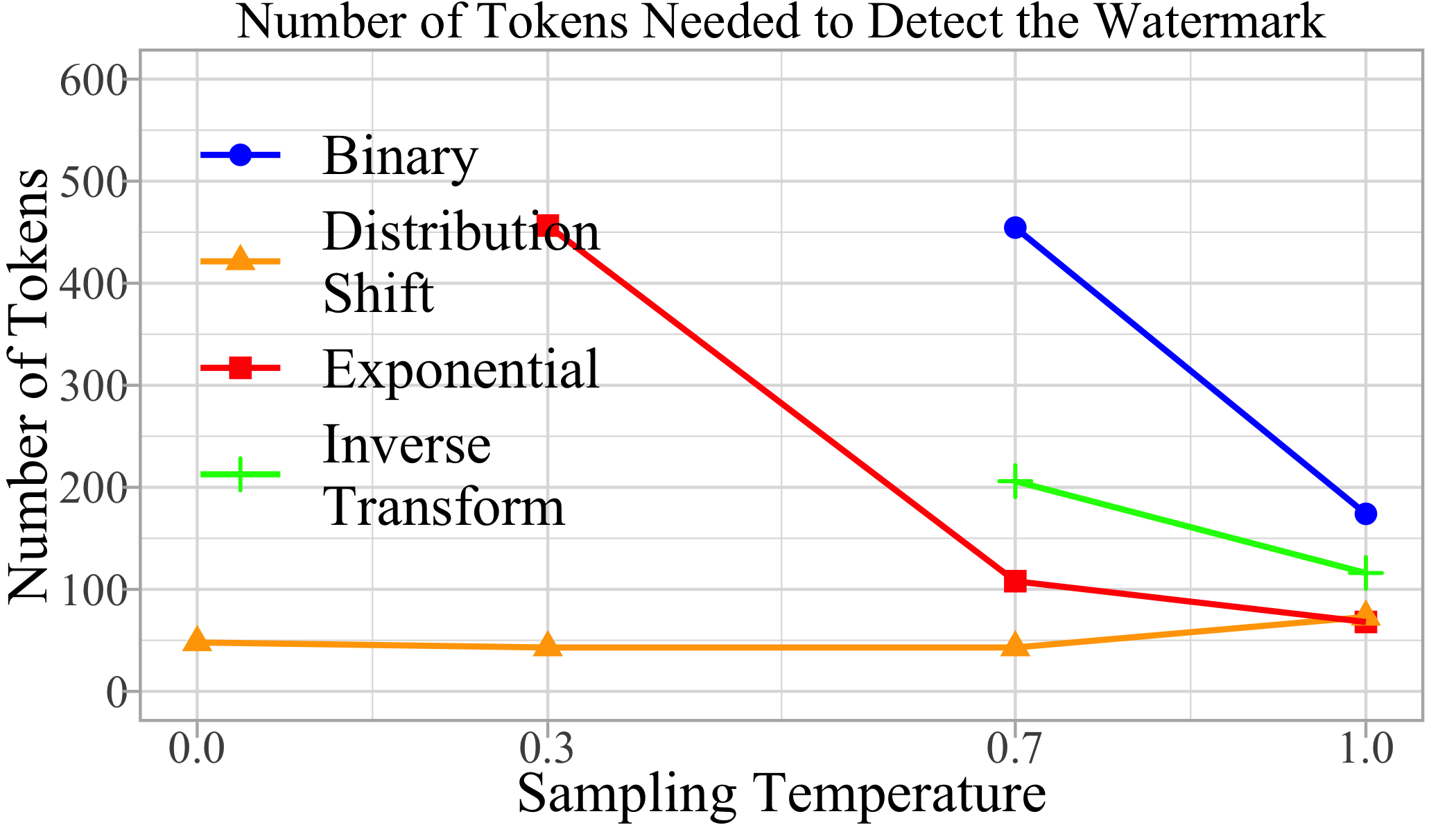}
    \end{minipage}%
        \hfill
    \begin{minipage}{0.38\textwidth}
        \caption{Watermark size at near-optimal quality for each watermarking schemes taken from the literature, using Mistral-7b-Instruct, at various sampling temperatures. The distribution-shift scheme~\cite{kirchenbauer_watermark_2023} outperforms all others and needs less than 80 tokens to be detected.}
        \label{fig:aggregate-mistral}
    \end{minipage}
\end{figure*}

\smallskip\noindent\textbf{Mistral.} The distribution-shift scheme is still superior when using Mistral-7B-Instruct~\citep{jiang2023mistral}. Mistral can be watermarked in 75 tokens or less, regardless of temperature. 
A version of \cref{fig:aggregate} using Mistral is in \cref{app:ssec:additional_figures} (\cref{fig:aggregate-mistral}).

\smallskip\noindent{\bf Tunability.}
We plot all quality/size/tamper resistance tradeoffs attainable, at a fixed temperature ($T=1$), in~\cref{fig:all_params}.
Each data point represents a combination of a watermarking scheme and a parameter setting.
Distribution-shift is the most tunable scheme: by adjusting the bias parameter, one can sacrifice quality 
for watermark size, something that is not possible for other watermarks.
The exponential scheme shows significant spread in quality because some parameter settings do not 
provide enough entropy (e.g., setting the sliding window too small).
Below, we describe how these schemes can be tuned.

\subsection{Parameter tuning}
\label{ssec:param_tuning}

\begin{table}[t]
\centering
 \begin{tabular}{cccccc} 
    \textbf{Parameter} & \rot{\parbox{1cm}{\centering\textbf{Win Size}}} & \rot{\parbox{1cm}{\centering\textbf{Min Hash}}} & \rot{\parbox{1cm}{\centering\textbf{Skip Prob}}} & \rot{\parbox{1cm}{\textbf{Bias}}} & \rot{\parbox{1cm}{\textbf{Key Len}}} \\
    \midrule
    Quality & + & $\perp$ & + & \textcolor{red}{\bf{-}} & + \\
    Size & + & + & + & \textcolor{red}{\bf{-}} & + \\
    \parbox{1.5cm}{\centering Tamper\\ Resistance} & \textcolor{red}{\bf{-}} & \textcolor{red}{\bf{+}} & $\perp$ & \textcolor{red}{\bf{+}} & \textcolor{red}{\bf{-}} \\
    \midrule
    \textbf{Suggestion} & 1-3 & False & 0.05 & $\geq$2 & 4 \\
    \bottomrule
    \end{tabular}
    \caption{Correlations between parameters and metrics. ``+'' indicates a positive correlation, ``-'' a negative correlation, ``$\perp$'' no correlation. 
    Symbols in red indicate a strong effect of the parameter on the given metric.}
    \label{table:parameter-effects}
\end{table}

Table~\ref{table:parameter-effects} summarizes the effects of each parameter on the metrics. The 
distribution-shift scheme is the most tunable: the bias $\delta$ allows to adjust the quality to watermarking size trade-off. We plot all quality/size/tamper resistance tradeoffs attainable, at a fixed temperature ($T=1$), in~\cref{fig:all_params}.

\smallskip\noindent{\bf Window size.} 
Increasing the window size increases quality, but also increases the watermark size and decreases robustness.
We recommend using a window size of 1 to 3; larger window sizes do not further improve quality.

\smallskip\noindent{\bf Min hash.} 
At a window size of 3, the min hash increases both tamper resistance and size by 33\%, compared to a sliding window.
We recommend using a simple sliding window.

\smallskip\noindent{\bf Skip probability.} 
We recommending setting the skip probability at 0.05 
for indistinguishable schemes (exponential, binary, inverse transform) 
using text-dependent randomness, 
as this adds non determinism to their outputs. 
This also provides slightly better quality than a skip probability of 0.

\begin{figure*}[h]
    \begin{minipage}{0.38\textwidth}
        \caption{Size and quality for varying biases, at $T=0.3$ and $T=1$. The quality is relative to the quality of the non-watermarked model at the given temperature. Increasing bias decreases size but also quality. Low temperature settings have less quality degradation.}
        \label{bias-fig}
    \end{minipage}
    \hfill
    \begin{minipage}{0.6\textwidth}
        \centering
        \includegraphics[width=\linewidth]{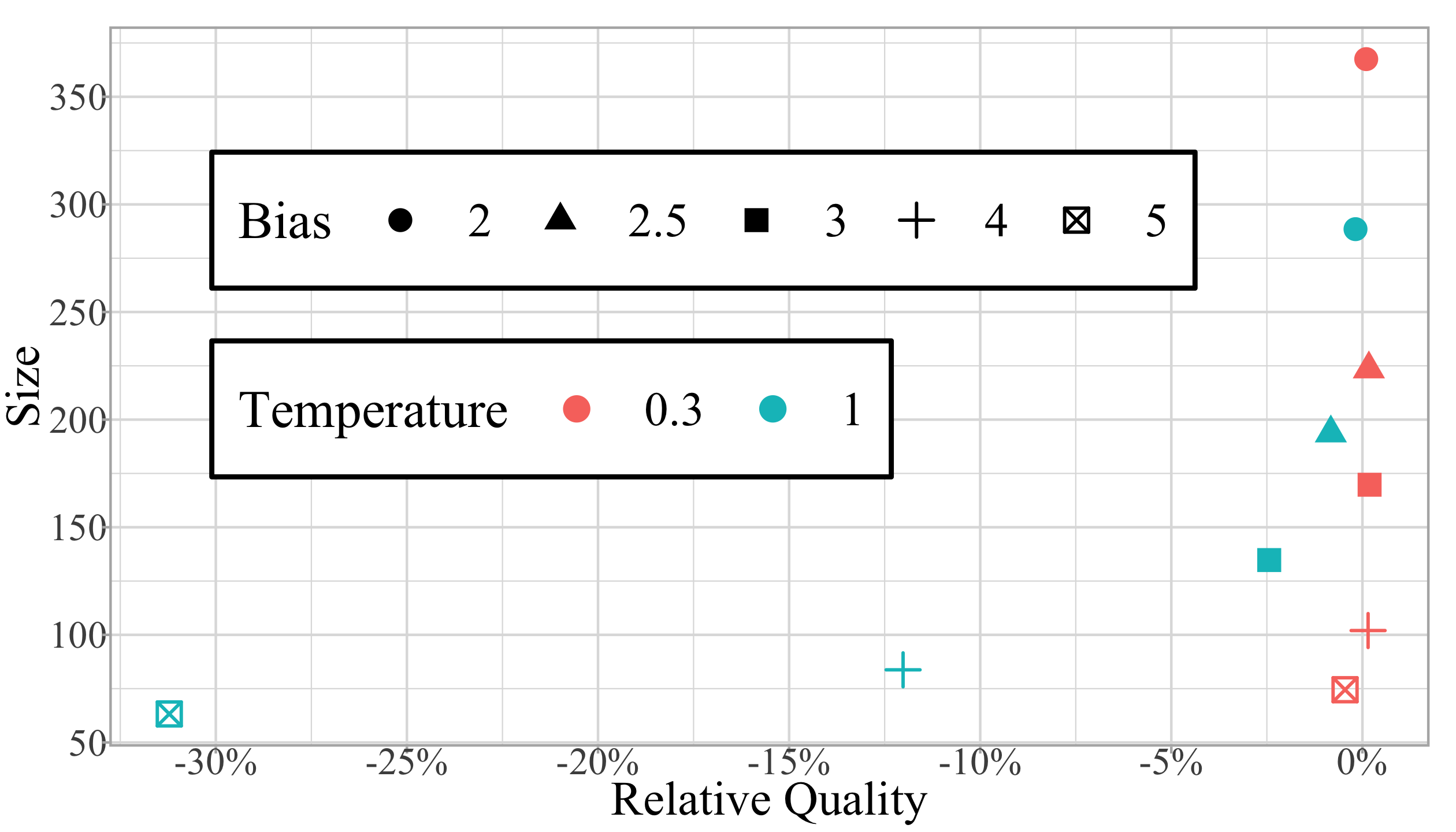}
    \end{minipage}%
\end{figure*}

\smallskip\noindent\textbf{Bias.} 
For distribution-shift, the bias ($\delta$) is a critical parameter
that has a large impact on quality, size, and tamper resistance.
\cref{bias-fig} visualizes its impact.
We suggest choosing $\delta$ based on the most common
temperature that the model is likely to be used with.
In all cases, only values of $\delta \geq 2$ yield efficient schemes.

\smallskip\noindent{\bf Green list size.}
Our early experiments showed that changing the green list size ($\gamma$) from 
0.5 only negatively impacts the watermark in all three metrics.
Therefore we fixed $\gamma$ to 0.5 for our experiments and suggest practitioners do the same. 

\begin{figure}
    \includegraphics[width=\linewidth]{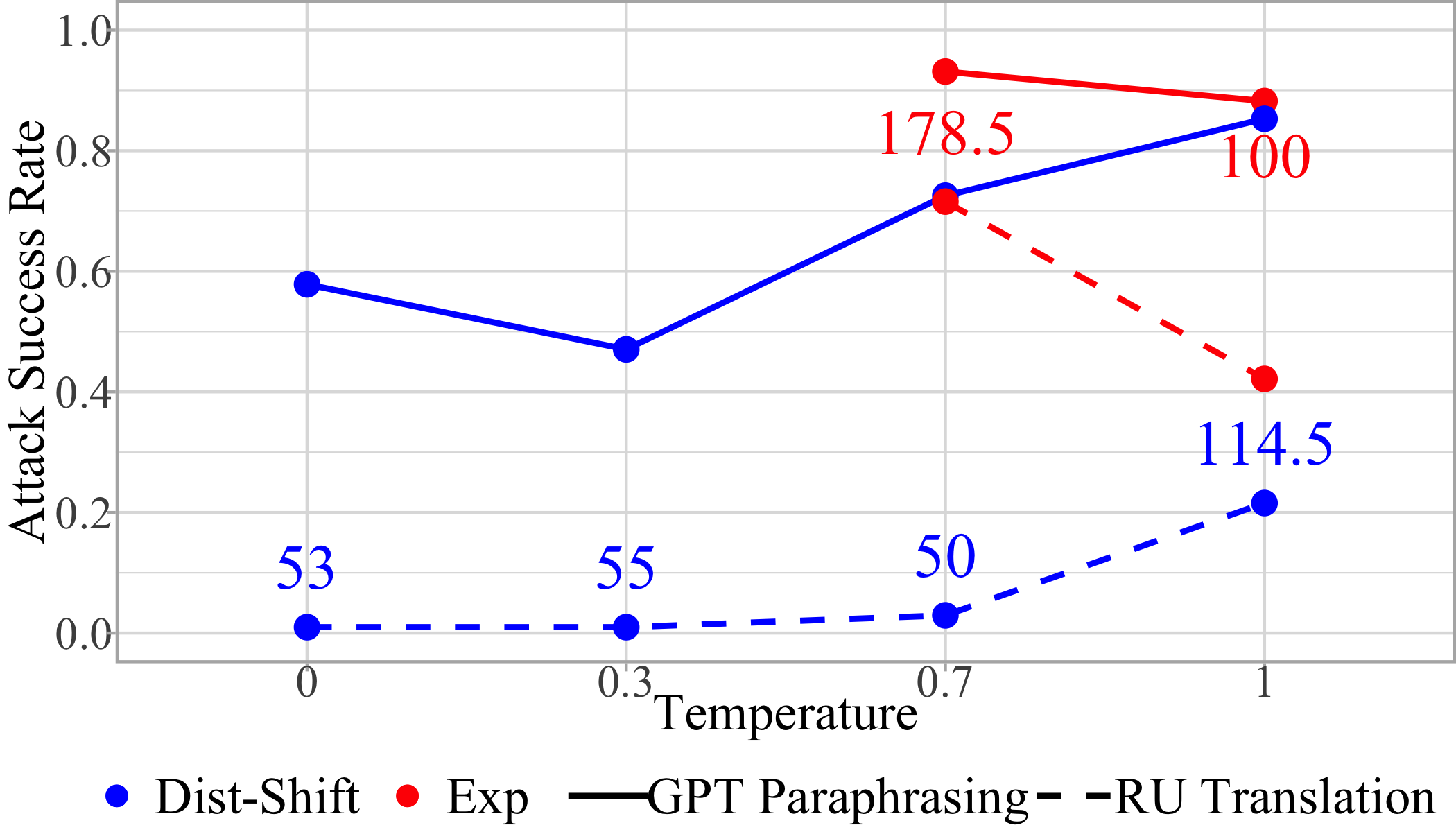}
    \centering
    \caption{Attack efficacy against top two schemes (with parameters from~\cref{fig:aggregate}). 
    The watermark size $E$ is labeled on each point.}
    \label{robustness-per-scheme}
\end{figure}

\begin{figure*}[h]
    \begin{minipage}{0.6\textwidth}
        \centering
        \includegraphics[width=\linewidth]{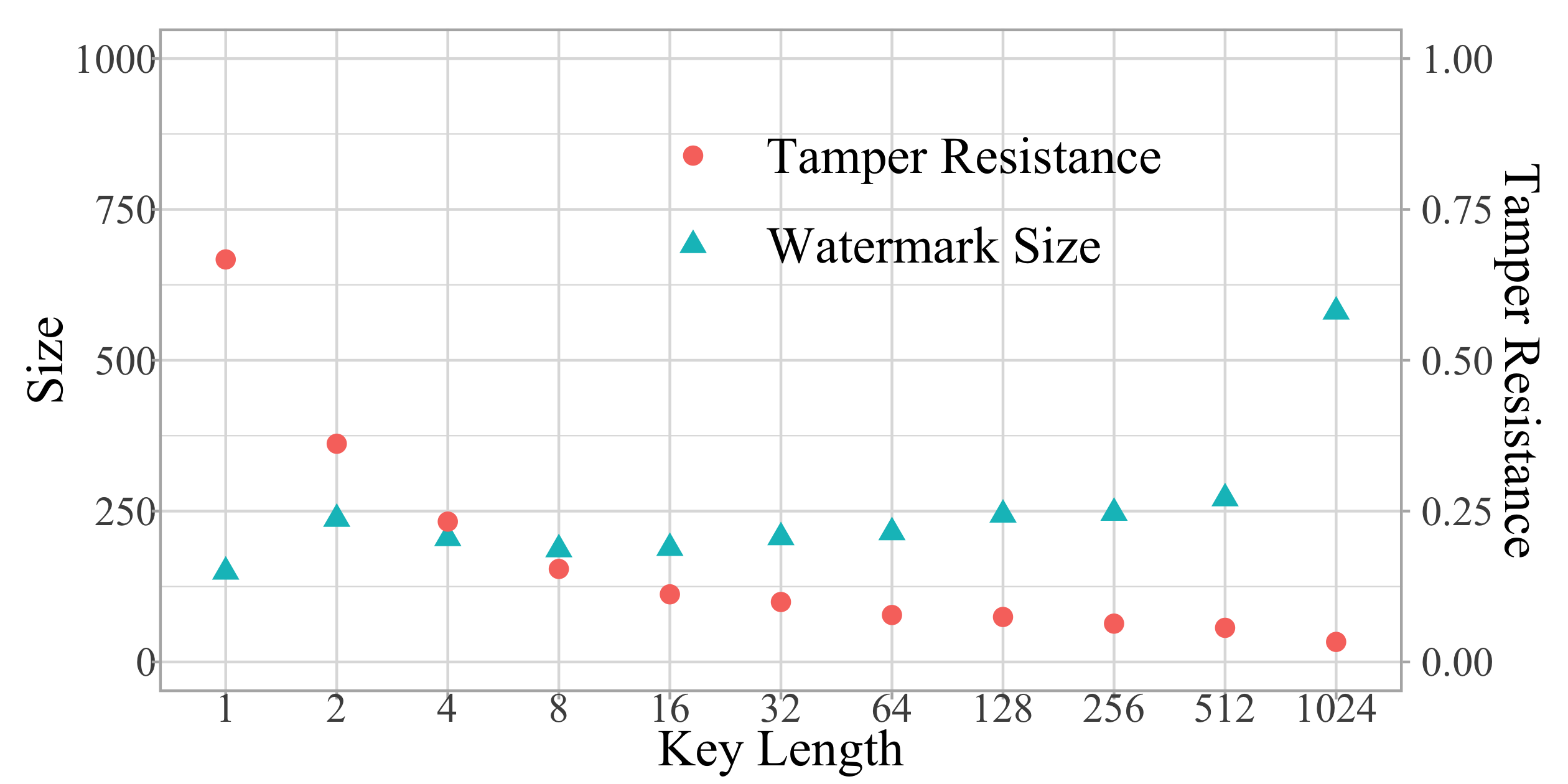}
    \end{minipage}%
    \hfill
    \begin{minipage}{0.38\textwidth}
        \caption{Size and tamper resistance as a function of key lengths (only using distribution-shift schemes with $\delta \leq 2$). Increasing key length increases size and decreases tamper resistance.}
        \label{keylen-fig}
    \end{minipage}
\end{figure*}

\smallskip\noindent{\bf Fixed randomness.} 
We recommend using a key length $L$ of 4.
Smaller values are detrimental to quality, but for $L \geq 4$ quality remains mostly the same, while larger values are worse for tamper resistance and size (\cref{keylen-fig}).

\subsection{Tamper resistance}
\label{ssec:robustness-eval}

\begin{figure}[t]
\centering
\includegraphics[width=\linewidth]{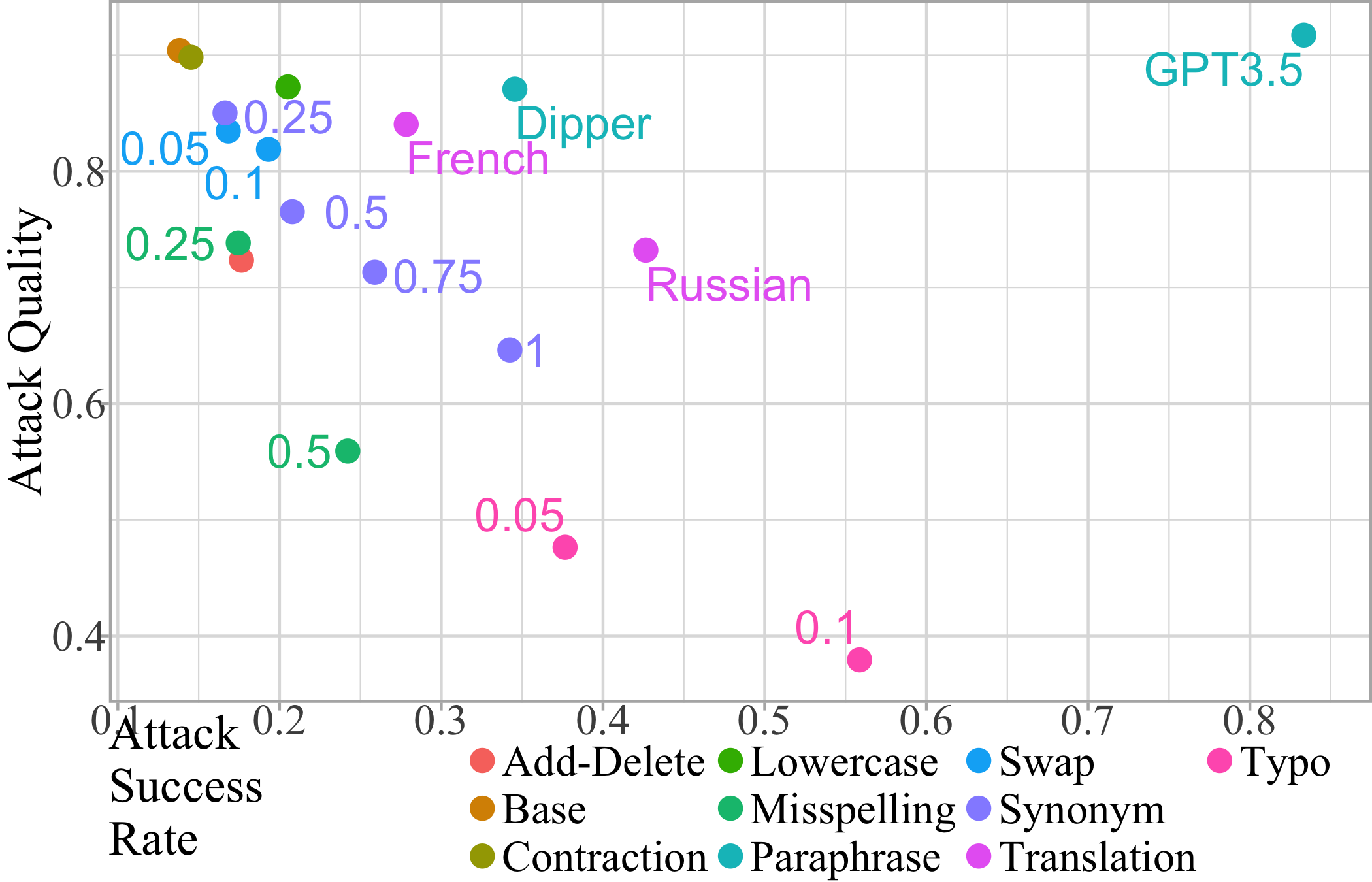}
\centering
\caption{Attack success rate and the relative generation quality after perturbation. Points are labeled with the attack's parameters, and have the same color for an attack class.}
\label{robustness-all-attacks}
\end{figure}

\noindent\textbf{Attack strength.}
We analyze the tamper resistance of schemes with the optimal parameters  
from~\cref{fig:aggregate}, evaluating the success rate of each
individual perturbation.
\cref{robustness-all-attacks} reports the average quality and success rate of perturbations.
Paraphrasing and translation are the most effective:
they remove the watermark most often, and do not heavily affect quality.
Using Dipper~\cite{krishna_paraphrasing_2023} to paraphrase is only slightly more effective than translation attacks, which are easier to run (\cref{ssec:tamper-resistance}).

\smallskip\noindent\textbf{Paraphrasing and translation attacks.} 
\cref{robustness-per-scheme} shows the success rate of GPT-3.5 paraphrasing
and Russian translation against the two best watermarking schemes 
(distribution shift and exponential, with optimal parameters from ~\cref{fig:aggregate}).
GPT-3.5 paraphrasing successfully removes the watermark at least 60\% of the time.
This success rate is unsurprising, since 
our parameters were chosen primarily to minimize size, not tamper resistance.
Russian translation is less successful, with under 20\% success rate at 
all but one temperature.
The success rates of both attacks are correlated: settings that are robust against one are 
also against the other.

\subsection{Metric variance}
\label{app:ssec:variance}

We added error bars to our main result (\cref{fig:aggregate}) by computing the 
median absolute deviation of watermark sizes when varying the random seed. These 
show that the gap in sizes between different schemes are statistically significant. 

The empirical variance of the LLM-rating quality metric remains mostly constant 
across different hyper-parameter choices, and decreases with temperature for all
schemes but distribution shift. In particular, for Llama 2, the baseline quality has 
a 95\% confidence interval of $0.90 \pm 0.005$. 

Tamper-resistance variance does change depending on the hyper-parameters. 
Schemes with large sizes tend to have more variance. Computing variance for all 
hyper-parameter choices would be too costly, instead, we computed empirical 
variance over a reduced set of hyper-parameters, and found an upper bound 
95\% confidence interval of $\pm 0.07$. 

The thresholds we select for quality and tamper-resistance metrics are larger 
than the typical the confidence interval to avoid excessive bias. 

\section{Discussion}
\label{sec:discussion}




\subsection{Indistinguishability}
\label{ssec:discussion:indistinguishability}
A watermark is indistinguishable if an adversary without the secret key cannot tell apart outputs from the watermarked and non-watermarked distributions of the same model. All but the distribution-shift schemes are indistinguishable from the original distribution when using appropriate randomness (fixed randomness or text-dependent randomness with long context windows).
Indistinguishability guarantees that the watermark does not affect the quality of the 
model.
Although useful to prevent model theft~\citep{he2022cater} and for code watermarking, we argue that indistinguishability is not crucial for watermarking natural language generations.

\smallskip\noindent\textbf{Indistinguishability for quality.} The distribution-shift scheme alters next-token distributions, so it could be detected given enough output tokens. Despite this, 
it does not degrade output quality with the right parameters, while requiring less tokens 
to be detected than all other watermarks. 

\smallskip\noindent\textbf{Indistinguishability for robustness.} Indistinguishability does not 
guarantee higher tamper resistance, as shown by the distribution-shift scheme's performance in~\cref{robustness-per-scheme}). In principle, distinguishability could help attackers 
identify and thus remove watermarks. However, much simpler attacks like paraphrasing are 
strong enough to remove many watermarks without requiring a distinguisher. 

We suspect the distribution-shift scheme performs better because
(1) it has the freedom to change the output distribution which offers more possibilities 
to embed a watermark; and (2) it works with the unmodified logits, before temperature scaling, 
so it can watermark text with low temperatures. 

\smallskip\noindent\textbf{Indistinguishability for structured tasks.}
Optimal watermarks for text do not work well on code (\cref{sec:results}). In fact, only generations from indistinguishable schemes with fixed randomness and key length $L > 16$ pass as many APPS tests as the baseline (in a 95\% confidence interval of $\pm$ 2.2\%). Of these schemes, the most efficient (exponential, $L=16$) only watermarks 28\% of output programs at $T=1$, doubles the watermark size of the best schemes for natural language, and fails to watermark for $T<1$. In a similar manner, distribution-shift watermarks decrease the accuracy of multiple choice queries. The introduced logit bias leads the LLM to select the wrong answer in cases where it is unsure. This is less pronounced for Mistral, which is on average more certain of its decisions. Both these tasks are sensitive to any token change, thus  indistinguishability is crucial for correctness. 

\subsection{Computational efficiency}
\begin{figure}
    \centering
    \includegraphics[width=\linewidth]{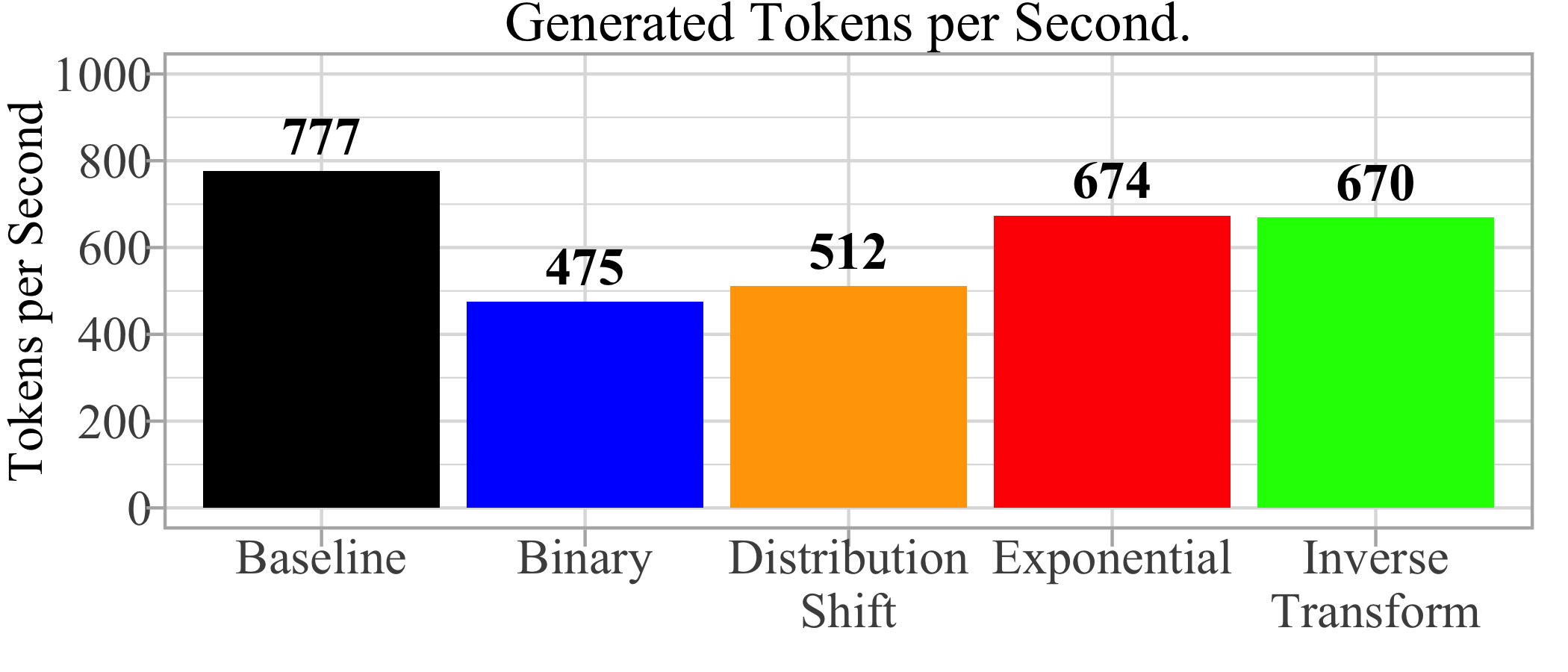}
    \caption{Generated tokens per second for each of the four sampling strategies.}
    \label{fig:tokenspersecond}
\end{figure}

The watermarking schemes we evaluate should not appreciably affect the time or cost to generate outputs from a LLM, since their sampling strategies have negligible complexity compared to the cost of running a transformer model. Detection algorithms are also much faster than LLM inference. Disparities in run-times are due to our unoptimized implementations. We show in~\cref{fig:tokenspersecond} the effect of the four sampling strategies on the number of generated tokens per second.

\subsection{Limitations}
\label{ssec:discussion:limitations}

\smallskip\noindent\textbf{Watermarking code.}
Our benchmark is focused on natural language tasks. More work is needed to better understand the impact of watermarking schemes on code-generation models, and to design schemes that work both for natural language and code tasks.

\smallskip\noindent\textbf{Benchmark size and coverage.}
Our benchmark only uses three language tasks and generates a total of 296 outputs for selecting optimal watermarking parameters. We include many more tasks for validation, however using more tasks and generations for the parameter selection process could help further evaluate watermarks, at the cost of increasing benchmark runtime. We only ran the benchmark on two open-sourced models, Llama2-7b and Mistral-7b. MarkMyWords is designed to compare watermarking techniques rather than specific LLMs. We found the key factor in watermarking to be the entropy in the model's next token probabilities, which does not necessarily correlate with model size or internal architecture. For instance, Mistral exhibits more entropy than Llama2 despite having the same number of parameters. We believe our conclusions extend to larger models.

\smallskip\noindent\textbf{Tamper resistance vs. robustness.} All watermarks can be broken by a sophisticated 
attacker~\cite{zhang2023watermarks}. We focus on evaluating {\it tamper resistance}, which is the 
watermark's ability to withstand output perturbations. We do not evaluate the attacks 
listed in~\cref{ssec:oos}, 
nor do we evaluate perturbations that require using another language model (its own outputs could also be watermarked).
We recognize our tamper-resistance 
metric is an upper bound on robustness against real-world sophisticated attackers.
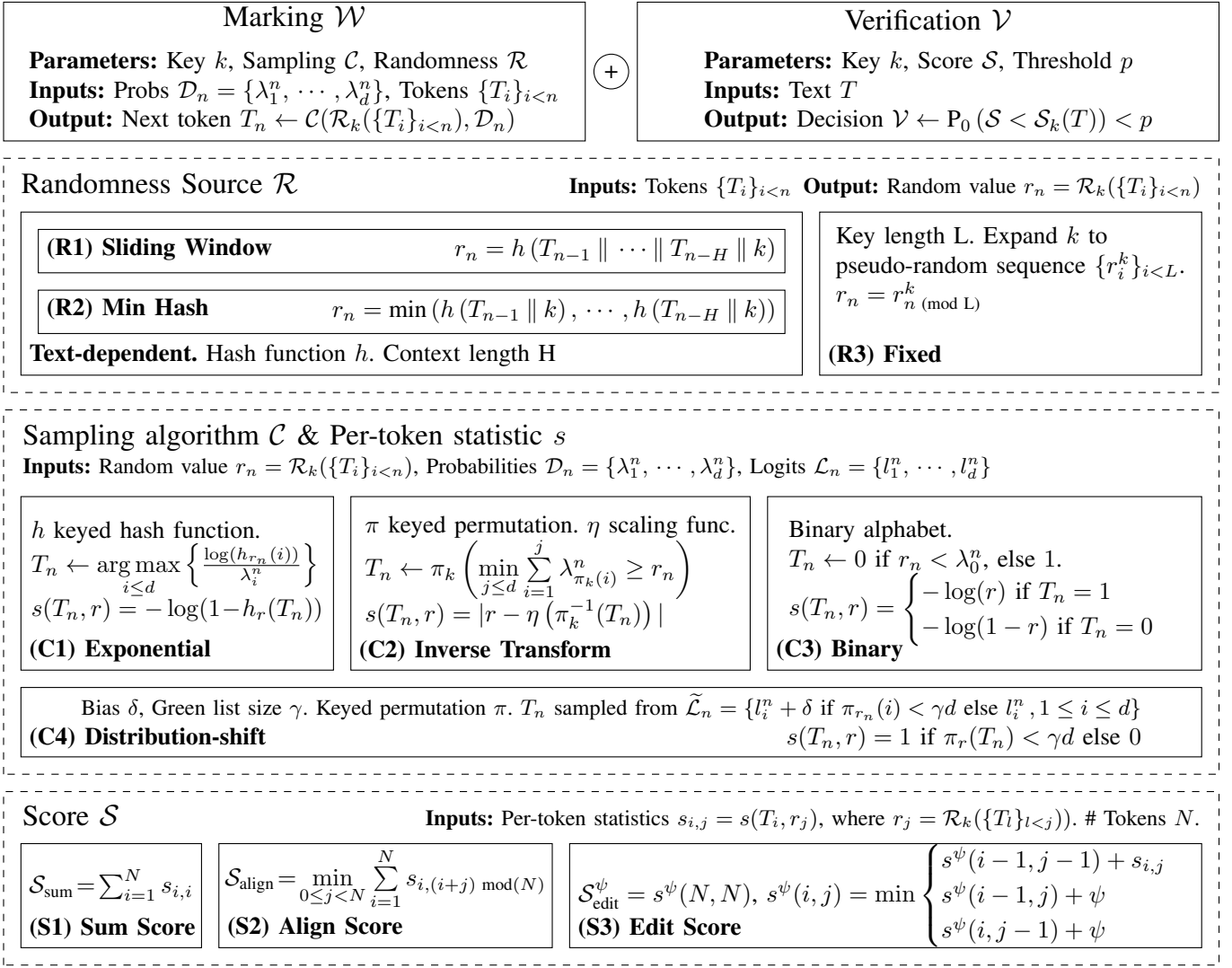
\begin{figure*}[h]
    \begin{center}   
    \resizebox{\linewidth}{!}{
    \begin{tikzpicture}

    \draw[draw=black] (0,11.875) rectangle ++(8.375,2) node[pos=0.5, align=left] 
    {\\
    \\
    \textbf{Parameters:} Key $k$, Sampling $\mathcal{C}$, Randomness $\mathcal{R}$\\
    \textbf{Inputs:} Probs $\mathcal{D}_n = \{\lambda^n_1,\, \cdots, \lambda^n_d\}$, Tokens $\{T_i\}_{i < n}$\\
    \textbf{Output:} Next token 
    $T_n \leftarrow \mathcal{C}(\mathcal{R}_k( \{T_i\}_{i < n}), \mathcal{D}_n)$};
    \draw[draw=black] (9.125,11.875) rectangle ++(8.375,2) node[pos=0.5, align=left] 
    {\\
    \\
    \textbf{Parameters:} Key $k$, Score $\mathcal{S}$, Threshold $p$\\
    \textbf{Inputs:} Text $T$\\
    \textbf{Output:} Decision $\mathcal{V} \leftarrow \text{P}_{0}\left( \mathcal{S} < \mathcal{S}_k(T)\right) < p$};
    \draw (8.75,12.875) circle (0.25) node {+};
    \draw[draw=none] (0,13.375) rectangle ++(8.375,.5) node[pos=0.5, align=left] {\large{Marking $\mathcal{W}$}};
    \draw[draw=none] (9.125,13.375) rectangle ++(8.375,.5) node[pos=0.5, align=left] {\large{Verification $\mathcal{V}$}};
    
    
    \draw[draw=black,dashed] (0,8.25) rectangle ++(17.5,3.375);
    \draw[draw=none] (0,10.725) rectangle ++(17.5,1) node[pos=0.5, align=center] {\large{Randomness Source $\mathcal{R}$} \hspace{3.5cm} \small{
    \textbf{Inputs:} Tokens $\{T_i\}_{i < n}$\,
    \textbf{Output:} Random value $r_n = \mathcal{R}_k(\{T_i\}_{i < n})$}};
    \draw[draw=black] (0.25,8.5) rectangle ++(11.25,2.35) node[pos=0, anchor=south west] {\textbf{Text-dependent.} Hash function $h$. Context length H};
    \draw[draw=black] (0.5,9.1) rectangle ++(10.75,0.625) node[anchor=north west] at (0.5, 9.725) {\textbf{(R2) Min Hash}} node[pos=1, anchor=north east, align=left] {
    $r_n = \text{min} \left( h\left( T_{n-1} \mathbin\Vert k\right), \, \cdots, h\left( T_{n-H} \mathbin\Vert k\right) \right)$\\
    };
    \draw[draw=black] (0.5,9.975) rectangle ++(10.75,0.625) node[anchor=north west] at (0.5, 10.6) {\textbf{(R1) Sliding Window}} node[pos=1, anchor=north east, align=left] {
    $r_n = h\left( T_{n-1} \mathbin\Vert \, \cdots \mathbin\Vert T_{n-H} \mathbin\Vert k\right)$\\
    };
    \draw[draw=black] (11.75,8.5) rectangle ++(5.5,2.35) node[pos=0, anchor=south west] {\textbf{(R3) Fixed}} node[pos=0.5, align=left] {Key length L. Expand $k$ to\\ pseudo-random sequence $\{r^k_i\}_{i<L}$.\\ 
    $r_n = r^k_{n \text{ (mod L)}}$ \\ \\ };
    
    
    \draw[draw=black,dashed] (0,2.75) rectangle ++(17.5,5.25);
    \draw[draw=none] (0,6.35) rectangle ++(17.5,1.75) node[pos=0.5, align=left] {\hspace{-3cm}\large{Sampling algorithm $\mathcal{C}$ \& Per-token statistic $s$}\\
    \hspace{-3cm}\small{\textbf{Inputs:} Random value $r_n = \mathcal{R}_k( \{T_i\}_{i < n})$, Probabilities $\mathcal{D}_n = \{\lambda^n_1,\, \cdots, \lambda^n_d\}$, Logits $\mathcal{L}_n = \{l^n_1,\,\cdots,l^n_d\}$} \\};

    \draw[draw=black] (11,4.25) rectangle ++(6.25,2.5) node[pos=0, anchor=south west] {\textbf{(C3) Binary}} node[pos=0.5, align=left] {Binary alphabet.\\ 
    $T_n \leftarrow 0$ if $r_n < \lambda^n_0$, else $1$. \\
    $s(T_n, r) = \begin{cases} -\log(r) \text{ if } T_n = 1\\
          -\log(1-r) \text{ if } T_n = 0\\\end{cases} $};
    
    \draw[draw=black] (5,4.25) rectangle ++(5.75,2.5) node[pos=0, anchor=south west] {\textbf{(C2) Inverse Transform}} node[pos=0.5, align=left] 
    {$\pi$ keyed permutation. $\eta$ scaling func.\\
    $T_n \leftarrow \pi_k \left( \min\limits_{ j \leq d } \sum\limits_{i=1}^j \lambda^n_{\pi_k (i)} \geq r_n \right)$ \\
    $s(T_n, r) = | r - \eta \left( \pi^{-1}_k(T_n) \right) | $\\};
    
    \draw[draw=black] (0.25,4.25) rectangle ++(4.5,2.5) node[pos=0, anchor=south west] {\textbf{(C1) Exponential}} node[pos=0.5, align=left] 
    {$h$ keyed hash function. \\
    $T_n \leftarrow \argmax\limits_{i \leq d} \left\{ \frac{\log \left( h_{r_n}\left( i \right) \right)}{\lambda^n_i} \right\}$ \\
    $s(T_n, r) = -\log(1 \! - \! h_r(T_n))$\\};

    \draw[draw=black] (0.25,3) rectangle ++(17,1) node[pos=0, anchor=south west] {\textbf{(C4) Distribution-shift}} node[pos=0.5, align=right] {\small Bias $\delta$, Green list size $\gamma$. Keyed permutation $\pi$. $T_n$ sampled from $\widetilde{\mathcal{L}}_n = \{l^n_i + \delta \text{ if } \pi_{r_n}(i) < \gamma d \text{ else } l^n_i\, , 1 \leq i \leq d\}$\\
    $s(T_n, r) = 1 \text{ if } \pi_{r}(T_n) < \gamma d \text{ else } 0$};
    
    
    \draw[draw=black,dashed] (0,0) rectangle ++(17.5,2.5);
    \draw[draw=none] (0,1.75) rectangle ++(17.5,0.75) node[pos=0.5, align=center] {\large{Score $\mathcal{S}$} \hspace{4cm} \small{
    \textbf{Inputs:} Per-token statistics $s_{i,j} = s(T_i, r_j)$, where $r_j = \mathcal{R}_k( \{T_l\}_{l < j}))$. \# Tokens $N$.}};

    \draw[draw=black] (8.15,0.25) rectangle ++(9.1,1.5) node[pos=0, anchor=south west] {\textbf{(S3) Edit Score}}
    
    node[pos=0.5, align=left] {
    $\mathcal{S}_{\text{edit}}^\psi = s^\psi(N,N)$,
    $
        s^\psi (i,j) = \min \begin{cases}
          s^\psi (i-1, j-1) + s_{i,j}\\
          s^\psi (i-1, j) + \psi\\
          s^\psi (i, j-1) + \psi\\
        \end{cases} 
    $};
    \draw[draw=black] (0.25,0.25) rectangle ++(2.6,1.5) node[pos=0, anchor=south west] {\textbf{(S1) Sum Score}} node[pos=0.5, align=left] {$\mathcal{S}_{\text{sum}}\! = \! \sum_{i=1}^N s_{i,i}$ \\};
    \draw[draw=black] (3.1,0.25) rectangle ++(4.8,1.5) node[pos=0, anchor=south west] {\textbf{(S2) Align Score}} node[pos=0.5, align=left] {$\mathcal{S}_{\text{align}} \!= \!\min\limits_{0 \leq j < N} \sum\limits_{i=1}^N s_{i, (i+j) \text{ mod}(N)}$ \\ \\ };
    
    \end{tikzpicture}}
    \caption{Watermarking design blocks. There are three main components: randomness source, sampling algorithm (and associated per-token statistics), and score function. Each solid box within each of these three components (dashed) denotes a design choice. The choice for each component is independent and offers different trade-offs.}\label{fig:design-figure}
    \end{center}
    \end{figure*}

\section{Designing a watermark}
\label{sec:taxonomy}

\ifappendixonly
\else
Following the overview in~\cref{sec:background}, we now detail the watermarking design space, introducing a taxonomy unifying previous schemes. First, we outline the requirements and building blocks for a text watermark, summarized in \cref{fig:design-figure}.
\fi

\subsection{Requirements}
A viable watermarking scheme must detect watermarked texts accurately without impairing the original model's utility. It should exhibit:

\smallskip\noindent \textbf{High recall:} Large $\Pr[\mathcal{V}_k(T) = \texttt{True}]$ for watermarked texts with key $k$.

\smallskip\noindent \textbf{High precision:} Large $\Pr[\mathcal{V}_k(\Tilde{T}) = \texttt{False}]$ for human-generated texts, regardless of key $k$.

\smallskip\noindent \textbf{Quality:} Comparable output quality to the original model.

\smallskip\noindent \textbf{Robustness:} Resistance to changes in watermarked texts.

\smallskip\noindent \textbf{Efficiency:} Low computational overhead to enable high-throughput verification by the LLM provider. 

\ifappendixonly
\else
\smallskip\noindent Additionally, desirable properties include diversity, enabling multiple outputs for a prompt (useful for beam-search generation), and \emph{undetectability} (\emph{indistinguishability}), wherein watermarked outputs are hard to distinguish from regular outputs (as discussed in~\cref{ssec:discussion:indistinguishability}).
\fi

\smallskip\noindent We focus on symmetric-key watermarking, where both the watermarking and verification procedures share a 
secret key. This is most suitable for proprietary language models served via an API.
Alternatively, one could publish the key, enabling anyone to run the verification procedure.

\subsection{Watermark design space}
\label{sec:watermark-design}

Designing a good watermark is a balancing act.
For instance, replacing every word of the output with \texttt{[WATERMARK]}
would achieve high recall but remove all the utility of the model.
%
Existing proposals have cleverly crafted marking procedures that 
are meant to preserve quality, provide high precision and recall, 
and achieve a degree of robustness.
Despite their apparent differences, we observe that they can all be expressed within a unified framework.

\smallskip\noindent The marking procedure $\mathcal{W}$ contains a randomness source $\mathcal{R}$ and a sampling algorithm $\mathcal{C}$.
    The randomness source $\mathcal{R}$ produces a (pseudorandom) value $r_n$ for each new token, based on the secret key $k$ and the previous tokens $T_0,\cdots,T_{n-1}$.
    The sampling algorithm $\mathcal{C}$ uses $r_n$ and the model's next token distribution $\mathcal{D}$ to select a token.
    
\smallskip\noindent The verification procedure $\mathcal{V}$ is a one-tailed significance test that computes a $p$-value for the null hypothesis that the text is not watermarked.
    The procedure compares this $p$-value to a threshold, setting the precision of the detector.
    In particular, we compute a per-token score $s_{n,m} \coloneqq s(T_n, r_m)$ for each token $T_n$ and randomness $r_m$, aggregate them to obtain an overall score $\mathcal{S}$, and compute a $p$-value from this score.
    We consider all overlaps $s_{n,m}$ instead of only $s_{n,n}$ to support scores that consider misaligned randomness and text after perturbation. 

Next, we show how each scheme we consider falls within this framework, each with its own choices for $\mathcal{R},\mathcal{C},\mathcal{S}$.

\subsection{Randomness source $\mathcal{R}$}\label{app:ssec:randomness}\label{app:ssec:binary}
\ifappendixonly
\else
As mentioned in~\cref{ssec:watermark-design}, randomness in watermarking can be generated in two primary manners: 
\fi
\emph{text-dependent} and \emph{fixed}. Both leverage a secret key to produce pseudorandom values, reproducible by the verification procedure. Text-dependent approaches, such as those by \citet{aaronson_watermarking_2022} and \citet{kirchenbauer_watermark_2023}, use prior tokens to generate randomness, relying on varied context lengths ($H$) and aggregation functions ($f$), including sliding window (\cref{fig:design-figure}, R1) and min hash (\cref{fig:design-figure}, R2): $r_n = f\left(T_{n-H},\,\cdots,T_{n-1},k\right)$, where $f := h\left( T_{n-1} \mathbin\Vert \, \cdots \mathbin\Vert T_{n-H} \mathbin\Vert k\right)$ for sliding window, and $f := \text{min} \left( h\left( T_{n-1} \mathbin\Vert k\right), \, \cdots, h\left( T_{n-H} \mathbin\Vert k\right) \right)$ for min hash. 

Fixed randomness (\cref{fig:design-figure}, R3), employed by \citet{kuditipudi_robust_2023}, generates values based on token index ($n$), using a repeated key sequence of length $L$ across generations: $r_n = f_k(n)$. 
When $L=1$ or $H=0$, both sources are identical, as $r_n$ will be the same value for 
every token. \citet{zhao2023provable} explored this option using the same sampling 
algorithm as~\citet{kirchenbauer_watermark_2023}. We analyze the impact of $H$ and $L$ 
in \cref{ssec:param_tuning}.

\citet{christ_undetectable_2023} set a target entropy for the context window 
instead of fixing a window size. This allows more precise control over the model's 
undetectability. However, one must try all context lengths to detect a 
watermark when using fixed entropy, thus we chose to keep using a fixed-size window for 
increased efficiency.

\subsection{Sampling algorithm $\mathcal{C}$}\label{app:ssec:sampling}
\ifappendixonly
\else
We now give more details about the four sampling algorithms initially discussed in~\cref{ssec:watermark-design}.
\fi

\smallskip\noindent\textbf{(\cref{fig:design-figure}, C1) Exponential.} Conceptualized by \citet{aaronson_watermarking_2022} and further employed by \citet{kuditipudi_robust_2023}, this algorithm leverages the Gumbel-max trick. Let $\mathcal{D}_n = \left\{\lambda^n_i\,, 1 \leq i \leq d\right\}$ be the distribution of the language model over the next token. 
The exponential scheme will select the next token as $T_{n} = \argmax\left\{ i \leq d,\,  \log \left( h_{r_n}\left( i \right) \right) / \lambda^n_i \right\}$ where $h$ is a keyed hash function using $r_n$ as its key.
The per-token variable used in the statistical test is either $s_n = h_{r_n}(T_n)$ or $s_n = -\log \left( 1-h_{r_n}(T_n)\right)$, yielding identical results. Prior work uses the latter quantity. We adhere to this for our benchmark, but analyze the former in~\cref{app:ssec:pseudorandom-proofs}.

\smallskip\noindent\textbf{(\cref{fig:design-figure}, C2) Inverse transform.} This scheme introduced by~\citet{kuditipudi_robust_2023} derives a random permutation using the secret key $\pi_k$. The next token is selected as the smallest index in the inverse permutation such that the CDF of the next token distribution is at least $r_n$. A detailed formula can be found in~\cref{fig:design-figure}. \citet{kuditipudi_robust_2023} propose using $s_n = | r_n - \eta \left( \pi^{-1}_k(T_n) \right) |$ as a the test variable, where $\eta$ normalizes the token index to the $[0,1]$ range.

\smallskip\noindent\textbf{(\cref{fig:design-figure}, C3) Binary.} Proposed by \citet{christ_undetectable_2023} for binary alphabets, this 
algorithm can adapt to any model by encoding tokens into binary sequences. In our implementation, 
we rely on a Huffman encoding of the token set, using frequencies derived from a large corpus of 
natural text. In this case, the distribution over the next token reduces to a single probability 
$p_n$ that token ``0'' is selected next, and $1-p$ that ``1'' is selected. The sampling rule selects
0 if $r_n < p$, and 1 otherwise. The test variable for this case is 
$s_n = -\log \left( T_n r_n + (1-T_n) (1-r_n) \right)$.

\smallskip\noindent\textbf{(\cref{fig:design-figure}, C4) Distribution-shift.} Suggested by \citet{kirchenbauer_watermark_2023}, 
it tweaks the token distribution to favor logits part of a green list. This list is selected using $r_n$ as a seed. The scheme adds bias $\delta$ to logits in the green list. Parameter $\gamma$ controls the size of the green list. 

The advantage of this last scheme over the others is that it preserves the model's diversity: 
for a given key, the model will still generate diverse outputs.
In contrast, for a given secret key and a given prompt, the first three sampling strategies 
will always produce the same result, since the randomness value $r_n$ will be the same.
\citet{kuditipudi_robust_2023} tackles this by randomly offsetting the key sequence of 
fixed randomness for each generation. We introduce a skip probability $p$ for the 
same effect on text-dependent randomness. Each token is selected without the marking 
strategy with probability $p$. We discuss generation diversity in~\cref{app:ssec:diverse}. 

Another advantage of the distribution-shift scheme is that it can also be used 
at any temperature, by applying the temperature scaling \emph{after} using the 
scheme to modify the logits. Other models apply temperature before watermarking.
The distribution-shift scheme is not indistinguishable from the original model. 
However, in practice, neither~\citet{aaronson_watermarking_2022} or~\citet{kuditipudi_robust_2023} 
are fully indistinguishable: \citet{gu2024learnability,jovanovic2024watermark} shows it is possible to learn a model that 
can spoof their watermarks.

\subsection{Score function $\mathcal{S}$}\label{app:ssec:score}
Determining the presence of a watermark in a text involves computing a score from per-token 
statistics. This score is then subject to a one-tailed statistical test where the null hypothesis 
is that the text is not watermarked. In other words, if its $p$-value is under a fixed threshold, 
the text is watermarked. Different works propose different scores.

\smallskip\noindent\textbf{(\cref{fig:design-figure}, S1) Sum score.}
\citet{aaronson_watermarking_2022} and \citet{kirchenbauer_watermark_2023} take the sum of all individual per-token statistics: $\mathcal{S}_{\text{sum}}=\sum_{i=1}^N s_i = \sum_{i=1}^N s(T_i, r_i).$
This assumes alignment between tokens $T_i$ and their corresponding random values $r_i$. Although effective for text-dependent randomness, it is not suited for fixed randomness. Any token removal at the text’s beginning, for instance, misaligns the subsequent $r_i$ values, compromising the watermark.

\smallskip\noindent\textbf{(\cref{fig:design-figure}, S2) Alignment score.}
Proposed by \citet{kuditipudi_robust_2023}, the alignment score aims to mitigate the misalignment issue.
Given the sequence of random values $r_i$ and the sequence of tokens $T_i$, the verification process now computes different versions of the per-token test statistic for each possible overlap of both sequences $s_{i,j} = s(T_i, r_j)$, and selects the minimum sum, as shown in~\cref{fig:design-figure}.

\smallskip\noindent\textbf{(\cref{fig:design-figure}, S3) Edit score.}
Similar to the alignment score, \citet{kuditipudi_robust_2023} propose the edit score as an alternative for dealing with the misalignment issue.
It comes with an additional parameter $\psi$ and is defined as $\mathcal{S}_{\text{edit}}^\psi = s^\psi(N,N)$, where $s^\psi(N,N)$ is defined as an edit score, detailed in~\cref{fig:design-figure}.

In all three cases, the average value of the score for watermarked text will be lower than for non-watermarked text.
%
In the case of the sum score, the previous works use the $z$-test on the score to determine whether the text is watermarked, but it is also possible, or even better in certain situations, to use a different statistical test according to \citet{fernandez_three_2023}.
When possible, we derive the exact distribution of the scores under the null hypothesis (\cref{app:ssec:exact_dist}) which is more precise than the $z$-test. When it is not, we rely on an empirical $T$-test, as proposed by \citet{kuditipudi_robust_2023}

\subsection{Score function considerations.}

\smallskip\noindent\textbf{Exact distribution of the score function.}\label{app:ssec:exact_dist}
The null hypothesis distribution for the exponential scheme with the regular test variable is an Irwin-Hall distribution centered with parameter $N$ (whose average quickly converges to a normal distribution centered in 0.5 with variance $\frac{1}{12N}$).
When using the $\log(\cdot)$ test variable, the null distribution is the Erlang distribution with parameter $N$. The binary scheme also follows an Erlang distribution, but with many more tokens since each token is broken down into a binary vector. The distribution-shift scheme has a null distribution a binomial with parameters $\gamma, N$. We derive these distributions in Appendix~\ref{app:ssec:pseudorandom-proofs}. However, for both other scores, and for the inverse transform, the null hypothesis distribution is too complex to compute. In these cases, verification uses a permutation test, as described in~\citet{kuditipudi_robust_2023}. Instead of comparing the score to a known distribution, we sample independent random sequences $\tilde{r}_i$ and compute the score of the text for that randomness: these trials are distributed like non watermarked text, so we can use them to compute an empirical p-value. 

\smallskip\noindent\textbf{Analysis of the edit score.} 
\label{app:ssec:editscore}
We analyzed the tamper resistance of the edit score on a subset of watermarks 
(distribution-shift with $\delta=2.5$ at a temperature of 1, for key lengths between 1 and 1024). 
We tried various $\psi$ values between 0 and 1 for the edit distance, and compared the tamper resistance 
and watermark size of the resulting verification procedures to the align score. 
Using an edit distance does improve tamper resistance for key lengths under 32, but at a large efficiency cost: 
for key lengths above 8, the edit score size is at least twice that of the align score. 
We do not recommend using an edit score on low entropy models such as Llama 2 chat.

\subsection{Limitations of building blocks}\label{app:ssec:limit_blocks}

Despite designing blocks for independence, certain scheme-parameter combinations are sub-optimal:

\smallskip\noindent\textbf{Sum score (S1)} lacks robustness with fixed randomness (R3).

\smallskip\noindent\textbf{Alignment score (S2)} is unsuitable for text-dependent randomness (R1, R2) since misalignment is not an issue.

\smallskip\noindent\textbf{Edit score (S3)} is only viable with text-dependent randomness (R1, R2) only for context length of 1. Beyond this, swapping, adding, or removing tokens affects random values rather than merely causing misalignment.

\noindent Furthermore, using context lengths of 0 or key lengths of 1 leads to having the same seed for each token. (S2) and (S3) are thus unnecessary since misalignment is not possible.

Our evaluation encompasses all logical combinations of randomness sources, sampling protocols, and verification scores along with their parameters. Due to the edit score's inefficiency, we primarily utilize sum and align scores. \cref{tab:design_space_combinations} lists the tested combinations. The distribution of non-watermarked scores is known for \textcolor{orange}{orange} configurations and unknown for \textcolor{blue}{blue} configurations. We rely on empirical $T$-tests~\cite{kuditipudi_robust_2023} for blue configurations. This method aims to benchmark against prior analyses and to explore under-researched combinations, potentially identifying superior configurations.

\begin{table}[t!]
    \centering
    \small
    \begin{tabular}{ccccc}
    \toprule
    & \textbf{C4} & \textbf{C1} & \textbf{C2} & \textbf{C3} \\
    & \makecell[c]{Distribution\\Shift} & Exponential & Binary & \makecell[c]{Inverse\\Transform} \\
    \midrule
    \textbf{S1+R1} & \textcolor{orange}{X} & \textcolor{orange}{X} & \textcolor{orange}{X} & \textcolor{blue}{X} \\
    \textbf{S1+R2} & \textcolor{orange}{X} & \textcolor{orange}{X} & \textcolor{orange}{X} & \textcolor{blue}{X} \\
    \textbf{S2+R3} & \textcolor{blue}{X} & \textcolor{blue}{X} & \textcolor{blue}{X} & \textcolor{blue}{X} \\
    \textbf{S3+R3} &  \textcolor{blue}{X} \\
    \bottomrule 
    \end{tabular}
    \caption{Tested combinations in the design space, using notations from~\cref{fig:design-figure}.
    }
    \label{tab:design_space_combinations}
\end{table}

\subsection{Techniques to enable diverse generations}\label{app:ssec:diverse}

For a fixed randomness source, \citet{kuditipudi_robust_2023} proposes to randomly shift the sequence of random values $\{r^k_i\}$ by an offset $s$, such that $r_n = r^k_{(s+n) \text{ (mod L)}}$. This means there are a total of $L$ possible unique values for $r_n$ depending on $s$. For a text-dependent randomness source, this trick does not work. 

Instead, one natural strategy is to randomly skip the watermarking selection procedure for some tokens, and instead sample the next token from the original multinomial distribution. We denote \textbf{S} this skip probability. 

Another strategy, discussed by \citet{christ_undetectable_2023}, is to only start watermarking text after enough empirical entropy has been generated: the first tokens are selected without a watermark. This accomplishes the same effect, and guarantees undetectability. However, as discussed in their Appendix, a user not wanting to generate watermarked text can simply run the model, keep the first few tokens, add them to the prompt, and start again. After repeating this step enough time, they can generate arbitrarily long text without a watermark. This seems like a larger practical drawback than loosing the guarantee of undetectability, thus we use the skip probability instead for promoting diversity. 
\section{Summary}

Our empirical analysis demonstrates existing watermarking schemes are nearly ready for deployment, 
providing effective methods to detect machine-generated text. Existing schemes can watermark Llama 2 with 
minimal quality loss in under 100 tokens, but still struggle at code generation.
We provide \benchmarkname{}, a benchmark to compare existing
and future LLM watermarking schemes.
We release our code in the hope it facilitates evaluation of watermarking
schemes and advances the discussion on
the desirable properties of watermarking schemes.

\bibliographystyle{plainnat}
\bibliography{refs}

\clearpage
\appendices
\section{Benchmark details}

\label{s:reprod-statement}
\label{app:ssec:prompts}
Our benchmark~(\cref{sec:experiments}) relies on a set of three prompt templates for each task, on three additional tasks for measuring watermarks in different situations, on a rating prompt, and a set of attacks. We provide details about the benchmark's implementation and the prompts we used in this section.

\subsection{Task prompts}

\vspace{1pt}\noindent\textbf{(1) Book report prompt.} ``Write a book report about X, written by Y.'', where X is a book title and Y is the book's author.

\vspace{1pt}\noindent\textbf{(2) Story generation prompt.} ``Write a X story about Y.'', where X is the tone, and Y is the topic.

\vspace{1pt}\noindent\textbf{(3) Fake news prompt.} ``Write a news article about X's visit to Y in Z.'', where X and Y are two political figures, and Z is a location.

\subsection{Validation tasks}

\vspace{1pt}\noindent\textbf{(V1) Legal Research.} ``Conduct legal research on X. Please summarize relevant regulations by region or country and include citations. Make conclusions based on your research about key developments in this field. Based on the current trends, make your forecast about how X will be regulated in the future.`` where X is the topic (e.g. AI)

\vspace{1pt}\noindent\textbf{(V1) RFC Summaries.} ``Write a detailed explanation and description of X: Y. Make sure is it self contained so I do not need to read the RFC myself to fully understand it.`` where X is the RFC number and Y the RFC title.

\vspace{1pt}\noindent\textbf{(V2) Multilingual.} ``Ecris un résumé detaillé de X, écrit par Y.``, where X is a book title and Y is the book's author.

\vspace{1pt}\noindent\textbf{(V3) Low entropy tasks.} We use 100 generations for task (1) at $T=1.0$ as the inputs for both paraphrasing and translation. Translation is done from English to French. 

\vspace{1pt}\noindent\textbf{(V4) Code generation.} We use 300 coding problems from the APPS dataset~\cite{hendrycks2021measuring}. We selected those with the longest median response, as to more effectively measure watermark size, and only chose problems with the lowest difficulty setting (``introductory''). 

\vspace{1pt}\noindent\textbf{(V5) Short summarization.} X is a news article, XX is a reference article and YY its summary (chosen to be the first element in the CNN/DM dataset~\cite{see2017get, hermann2015teaching})

\begin{small}
\begin{tcolorbox}[colback=black!5!white,colframe=black!75!white,left=0pt,right=0pt,top=0pt,bottom=0pt,breakable]
Extract a 3 sentence highlight from the following article. Here is one example for reference.\\
Example:\\
* Article: XX

* Highlights: YY

Current Task:\\
* Article: X

* Highlights:
\end{tcolorbox}
\end{small}

\vspace{1pt}\noindent\textbf{(V6) Question answering.} X is a question, Y are choices, XX is a reference question, YY its reference choices, and ZZ its answer (chosen to be the first element in the MMLU dataset~\cite{hendryckstest2021})

\begin{small}
\begin{tcolorbox}[colback=black!5!white,colframe=black!75!white,left=0pt,right=0pt,top=0pt,bottom=0pt,breakable]
Here is a question. You must select one of the options, and return the letter of the correct answer inside brackets. Here is an example for reference:\\

Question: XX

Choices: YY

Answer: ZZ\\

Now here's the current question:

Question: X

Choices: Y

Answer:
\end{tcolorbox}
\end{small}

\subsection{Rating prompt}

\begin{small}
\begin{tcolorbox}[colback=black!5!white,colframe=black!75!white,left=0pt,right=0pt,top=0pt,bottom=0pt,breakable]
<<SYS>> You are given a prompt and a response, and you provide a grade out of 100 measuring the quality of the response, in terms of accuracy, level of details, and typographical, grammatical and lexical correctness. Remove points as soon as one of the criteria is missed. <</SYS>>\\
Prompt: <TASK PROMPT>\\
Response: <MODEL OUTPUT>
\end{tcolorbox}
\end{small}
\section{Additional analysis}
\label{app:ssec:additional_figures}

Here we present additional figures to support results in the main text and discuss further findings.

\clearpage

\clearpage

\begin{table*}[t]
    \centering
    \small
    \begin{tabular}{z{2cm}lllll}
    \toprule
    \textbf{Constraint} & \textbf{Temp.} & \textbf{Distribution Shift} & \textbf{Exponential} & \textbf{Inverse Transform} & \textbf{Binary} \\
    \midrule
    \multirow{4}{*}{\parbox{2cm}{$1\%$ Quality, \\ $>0.2$ Tamper Resistance}} & 0 & 53 (1.5) & $\infty$ & $\infty$ & $\infty$ \\
    & 0.3 & 55 (0.5) & $\infty$ & $\infty$ & $\infty$ \\
    & 0.7 & 50 (1.0) & 178.5 (6.0) & 453.5 (15.0) & $\infty$ \\
    & 1 & 114.5 (2.5) & 100 (2.5) & 205 (4.0) & 372.5 (16.5) \\
    \midrule
    \multirow{4}{*}{\parbox{2cm}{$1\%$ Quality, \\ No Tamper Resistance}} 
    & 0 & 53 (1.5) & $\infty$ & $\infty$ & $\infty$ \\
    & 0.3 & 55 (0.5) & 980 (45) & $\infty$ & $\infty$ \\
    & 0.7 & 50 (1.0) & 178.5 (6.0) & 457.5 (10.0) & $\infty$ \\
    & 1 & 114.5 (2.5) & 100 (2.5) & 195 (7.0) & 372.5 (16.5) \\
    \midrule
    \multirow{4}{*}{\parbox{2cm}{$10\%$ Quality, \\ $>0.2$ Tamper Resistance}} 
    & 0 & 55 (1.0) & $\infty$ & $\infty$ & $\infty$ \\
    & 0.3 & 55 (0.5) & $\infty$ & $\infty$ & $\infty$ \\
    & 0.7 & 55 (1.0) & 168.5 (7.0) & 453.5 (15.0) & $\infty$ \\
    & 1 & 73 (1.5) & 80.5 (2.0) & 205 (4.0) & 372.5 (16.5) \\
    \midrule
    \multirow{4}{*}{\parbox{2cm}{$10\%$ Quality, \\ No Tamper Resistance}} 
    & 0 & 55 (1.0) & $\infty$ & $\infty$ & $\infty$ \\
    & 0.3 & 55 (0.5) & $\infty$ & $\infty$ & $\infty$ \\
    & 0.7 & 55 (1.0) & 168.5 (7.0) & 457.5 (10.0) & $\infty$ \\
    & 1 & 73 (1.5) & 80.5 (2.0) & 195 (7.0) & 372.5 (16.5)\\
    \bottomrule
    \end{tabular}
    \ifappendixonly
\else
    \caption{Size of ideal watermark under four tested constraints, for each sampling temperature and watermark on Llama 2. These are the values used in~\cref{fig:aggregate}. In parenthesis, the empirical median absolute deviation.}
    \fi
    \label{fig:detailed_llama_values}
\end{table*}

\begin{table*}[b]
    \centering
    \small
    \begin{tabular}{z{2cm}lllll}
    \toprule
    \textbf{Constraint} & \textbf{Temp.} & \textbf{Distribution Shift} & \textbf{Exponential} & \textbf{Inverse Transform} & \textbf{Binary} \\
    \midrule
    \multirow{4}{*}{\parbox{2cm}{$1\%$ Quality, \\ $>0.2$ Tamper Resistance}}
    & 0 & 48 & $\infty$ & $\infty$ & $\infty$ \\
    & 0.3 & 43 & 456.5 & $\infty$ & $\infty$ \\
    & 0.7 & 43 & 108 & 206 & 454.5 \\
    & 1 & 73 & 68 & 116 & 174 \\
    \midrule
    \multirow{4}{*}{\parbox{2cm}{$1\%$ Quality, \\ No Tamper Resistance}} 
    & 0 & 48 & $\infty$ & $\infty$ & $\infty$ \\
    & 0.3 & 43 & 456.5 & $\infty$ & $\infty$ \\
    & 0.7 & 43 & 108 & 224 & 454.5 \\
    & 1 & 73 & 68 & 116 & 174 \\
    \midrule
    \multirow{4}{*}{\parbox{2cm}{$10\%$ Quality, \\ $>0.2$ Tamper Resistance}} 
    & 0 & 45 & $\infty$ & $\infty$ & $\infty$ \\
    & 0.3 & 39.5 & 456.5 & $\infty$ & $\infty$ \\
    & 0.7 & 43 & 101 & 206 & 454.5 \\
    & 1 & 38 & 58 & 116 & 174 \\
    \midrule
    \multirow{4}{*}{\parbox{2cm}{$10\%$ Quality, \\ No Tamper Resistance}} 
    & 0 & 45 & $\infty$ & $\infty$ & $\infty$ \\
    & 0.3 & 39.5 & 456.5 & $\infty$ & $\infty$ \\
    & 0.7 & 43 & 101 & 224 & 454.5 \\
    & 1 & 38 & 58 & 116 & 174 \\
    \bottomrule
    \end{tabular}
    \caption{Size of ideal watermark under four tested thresholds, for each sampling temperature and watermark on Mistral. These are the values used in~\cref{fig:aggregate-mistral}.}
    \label{fig:detailed_mistral_values}
\end{table*}

\clearpage
\section{Exponential scheme proofs}
\label{app:ssec:pseudorandom-proofs}

We now analyze the exponential scheme when the statistical test used is $s_n = h_{r_n}(T_n)$.
We use the same notation as in~\cref{fig:design-figure}. In particular, $h_s$ is
a secure hash function mapping strings to $[0,1]$ with seed $s$, $k$ is a key selected uniformly 
at random amongst a set of keys $K$. Consider an execution of the language model which produced 
tokens $T_1,\ldots,T_n$ up until now (including the prompt). The next token distribution is represented as
$\mathcal{D}_{T_1,\cdots,T_n} = \left\{\lambda_i, P(T_{n+1} = i \mid T_1,\cdots, T_n) = \lambda_i\right\}$.
For an i.i.d randomness source (producing i.i.d random values), 
the exponential selection procedure has the following properties:

\begin{enumerate}[leftmargin=*,nosep]
    \item[\textbf{P1}] For a uniformly random key $k$, the distribution over the next token is the 
    same as without the watermark.
        \begin{align}
            P_k(\widetilde{T}_{n+1} = i \mid T_{i\leq n} ) 
            = P(T_{n+1} = i \mid T_{i\leq n} )
            = \lambda_i
        \end{align}
    \item[\textbf{P2}] The expectation of the hash of the next token is equal to the spike 
    entropy\footnote{Spike entropy is defined in~\cite{kirchenbauer_watermark_2023}. For a 
    discrete distribution $\mathcal{D} = \left\{ \lambda_1, \cdots, \lambda_n \right\}$, the 
    spike entropy $S\left(\mathcal{D}, t\right)$ (said \textit{modulus} i) is defined as 
    $\sum_{j=1}^{n} \frac{\lambda_j}{t+\lambda_j}$.} of the next token distribution. This is 
    always larger than the expectation of the hash of the next token without watermark, only 
    being equal for a degenerate distribution with only one token having a non-zero probability.
        \begin{align}
        \begin{split}
            &\frac{n}{n+1} \geq \mathbb{E}_k\left[ h_s\left(\widetilde{T}_{n+1}\right) \right] = S\left(\mathcal{D}_{T_1,\cdots,T_n},\, 1\right) \\
            &S\left(\mathcal{D}_{T_1,\cdots,T_n},\, 1\right) = \sum_{j=1}^{d} \frac{\lambda_j}{1+\lambda_j} \geq \mathbb{E}_k\left[ h_s\left(T_{n+1}\right) \right] = \frac{1}{2}
        \end{split}
        \end{align}
\end{enumerate}

\noindent
When using the sum score, we perform a hypothesis test, where under $H_0$ the text is not 
watermarked, and under $H_1$ it is. In particular, let 
$\bar{h} = \frac{1}{n} \sum_{i=1}^{n} h_s\left(T_{i}\right)$ be the average hash over a text of 
$n$ tokens. Let $S_i$ be the average spike entropy of modulus $i$ over the text. 
$\bar{h}$ follows a Bates distribution under $H_0$ with parameter $n$ (which is well approximated by 
a Gaussian with average $0.5$ and variance $\frac{1}{12n}$. 
The distribution under $H_1$ has an average of $S_1$ and asymptotically follows a Gaussian 
distribution centered in $S_1$ with variance $\frac{S_2 - S_1^2}{n}$. 

\paragraph{Proof of \textbf{P1}}

Let $r = \mathbf{R}(k,  \{T_i\}_{i < n})$ be our randomness source. We assume that it is uniformly distributed between 0 and 1, for $k \sim \mathcal{U}(K)$, with $|K|$ large enough. Since $h_s$ is a secure hash function, we posit: 
\begin{enumerate}[leftmargin=*,nosep]
    \item[\textbf{A1}] The hash of each token is uniformly distributed: $h_r(i) \sim \mathcal{U}\left([0,1]\right) ~\forall i, T_1, \cdots, T_n$.
    \item[\textbf{A2}] The token hashes $\{ h_r(i),\, 1 \leq i \leq d\}$ are mutually independent.
\end{enumerate}
We have:
\small
\begin{align}
    P_k(\widetilde{T}_{n+1} = i | T_{i\leq n}) = P_k\left\{ \forall j \neq i,\, \frac{\log(h_r(i))}{\lambda_i} > \frac{\log(h_r(j))}{\lambda_j}\right\}
\end{align}
\normalsize
We can simplify this expression: if $\{h_r(i)\}_i$ are mutually independent uniformly distributed 
random variables between 0 and 1, then $\left\{\frac{-\log(h_r(i))}{\lambda_i}\right\}_i$ are 
mutually independent exponentially distributed random variables, with parameters $\{\lambda_i\}_i$. 
By writing $u_i = -\frac{\log(h_r(i))}{\lambda_i}$, we then have:
\begin{align}
    P_k(\widetilde{T}_{n+1} = i | T_{i\leq n}) = P_k\left\{ u_i < \min_{j \neq i}(u_j) \right\}
\end{align}

We now use a useful property of exponential random variables: The minimum of a set of independent exponential random variables of parameters $\lambda_1, \cdots, \lambda_n$ is also an exponential random variable, with parameter $\sum_{i=1}^n \lambda_i$. Thus, $\min_{j \neq i}(u_j) \sim \mathrm{Exp}(1-\lambda_i)$ (since $\lambda_i$ are a probability distribution, they sum to 1, so $\sum_{j\neq i}\lambda_j = 1 - \lambda_i$). We can now finish the proof.
\begin{align}
    \begin{split}
        P_k(&\widetilde{T}_{n+1} = i | T_{i\leq n})
        = P_k\left\{ u_i < \min_{j \neq i}(u_j) \right\} \\
        &= \int_0^{\infty} \lambda_i e^{-\lambda_i x} \int_x^{\infty} (1-\lambda_i) e^{-(1-\lambda_i) y} dx dy\\
        &= \lambda_i = P(T_{n+1} = i | T_{i\leq n}).
    \end{split}
\end{align}

\paragraph{Proof of \textbf{P2}}

For a sequence of previous tokens $T_{i\leq n} = T_1,\cdots,T_n$, lets compute the expected value of the hash of the next token under both the watermarked and non-watermarked model.

As discussed above, for $k \sim \mathcal{U}(K)$, $h_r(i) \sim \mathcal{U}([0,1])$ and are mutually independent, and we have $\mathbb{E}_k\left[ h_r(i) \right] = \frac{1}{2}$

For the non-watermarked model, we select a token independently from the key $k$ or the values $h_r(i)$, thus:
\begin{align}
    \begin{split}
        \mathbb{E}_k\left[ h_r\left(T_{n+1}\right) \right]
        &= \mathbb{E}_k\left[ \sum_{i=1}^d \mathbbm{1}_{\{T_{n+1} = i\}} h_r\left(i\right) \right]\\
        &= \sum_{i=1}^d \mathbb{E}\left[ \mathbbm{1}_{\{T_{n+1} = i\}} \right]  \mathbb{E}_k\left[ h_r\left(i\right) \right]\\
        &= \sum_{i=1}^d \frac{1}{2}\lambda_i = \frac{1}{2}
    \end{split}
\end{align}

For the watermarked model, token selection is no longer independent from the key $k$ or 
the hash values. Instead, we use the notations from the previous proof to compute this 
expectation. In particular, we have 
$\{T_{n+1}=i\} = \left\{ \forall j \neq i,\, u_i < u_j \right\} = \left\{ u_i < \min_{j \neq i}(u_j) \right\} $, 
with $u_i$ exponentially distributed with parameter $\lambda_i$, and $\min_{j \neq i}(u_j)$ 
exponentially distributed with parameter $1-\lambda_i$, both independent. Also, 
since $u_i = -\frac{\log(h_r(i))}{\lambda_i}$, we have $h_r(i) = e^{-u_i \lambda_i}$.
\begin{align}
    \begin{split}
        \mathbb{E}_k&\left[ h_r\left(T_{n+1}\right) \right] = \mathbb{E}_k\left[ \sum_{i=1}^V \mathbbm{1}_{\{T_{n+1} = i\}} h_r\left(i\right) \right]\\
        &= \sum_{i=1}^d \iint_0^{\infty} e^{-\lambda_i x} \mathbbm{1}_{x < y} \lambda_i e^{-\lambda_i x} (1-\lambda_i) e^{-(1-\lambda_i) y} dx dy\\
        &= \sum_{i=1}^d \frac{\lambda_i}{1 + \lambda_i}
    \end{split}
\end{align}
Thus, the expectation of the watermarked model's next token hash is equal to the spike entropy of next token distribution (as defined in~\cite{kirchenbauer_watermark_2023}). Analysis of the spike entropy shows that its minimum value is $0.5$, when all but one token have 0 probability, and its maximum is $\frac{d}{d+1}$, when all tokens are equally probable.

\paragraph{Verification}
The verification procedure computes the average hash value over all tokens in a text: $\bar{h} = \frac{1}{n} \sum_{i=1}^{n} h_r\left(T_{i}\right)$.
Let's analyze this random variable $\bar{h}$ in both the watermarked and non-watermarked settings. 
In the regular case, we can show, for $k \sim \mathcal{U}(K)$, that $h_r\left(T_{i}\right) \sim \mathcal{U}([0,1])$, for some $x \in [0,1]$:
\begin{align}
\begin{split}
P_k&\left( h_r\left(T_{i}\right) < x \right) \\
&= \sum_{j=1}^d P_k\left( \{T_i = j\} \cap \{h_r\left(i\right) < x \}\right) \\
&= \sum_{j=1}^d P\left( \{T_i = j\} \right) P\left(\{h_r\left(i\right) < x\}\right) = x
\end{split}
\end{align}
Which is the CDF of a continuous uniform random variable between 0 and 1.
Furthermore, since the randomness source is i.i.d, the ${h_r(T_i)}_i$ are independent.
Thus $\bar{h}$ is the average of $n$ independent uniform random variables over 
$[0,1]$, so it follows a Bates distribution. When $n$ increases to $+\infty$, 
$\mathcal{B}(n)$ is equivalent to $\mathcal{N}(0.5, \frac{1}{12n})$. In practice, 
even for small values of $n$, the Bates distribution is close to Gaussian.

In the watermarked case, we start by computing the CDF and PDF of the hash of a single token, 
$h_r\left(T_{i}\right)$.
\small
\begin{align}
    \begin{split}
        &P_k\left( h_r\left(T_{i}\right) < x \right) = \sum_{j=1}^d P_k\left( \{T_i = j\} \cap \{h_r\left(j\right) < x \}\right) \\
        &= \sum_{j=1}^d \iint_0^{\infty} \mathbbm{1}_{z < y} \mathbbm{1}_{z > -\frac{\log(z)}{\lambda_j}} \lambda_j e^{-\lambda_j z} (1-\lambda_j) e^{-(1-\lambda_j) y} dz dy\\
        &= \sum_{j=1}^d \lambda_j x^{\frac{1}{\lambda_j}}.
    \end{split}
\end{align}
\normalsize
Thus the hash of a token has CDF $\sum_{j=1}^d \lambda_j x^{\frac{1}{\lambda_j}}$ and PDF $\sum_{j=1}^d x^{\frac{1}{\lambda_j} - 1}$. Using these formulas, we can derive higher order moments for the hash: $\mathbb{E}_k\left[ h_r\left(T_i\right)^{m}\right] = \sum_{j=1}^{d} \frac{\lambda_j}{1 + m\lambda_j}$. In particular, each moment of order $m$ is bounded between 1 and $\frac{V}{m+V}$. We denote $S_m$ to be the average moment of order $m$ over the text (which happens to also be the average spike entropy of modulus $m$).

We define $s^2_n = \sum_{i=1}^{n} \mathbb{E}_k\left[ \left(h_r\left(T_i\right) - 
\mathbb{E}_k\left[h_r\left(T_i\right)\right]\right)^2 \right] = n\left(S_2 - S_1^2\right)$.
Since we assume each hash is independent, thanks to the i.i.d randomness, we 
can use Lyapunov's central limit theorem on the sequence of hash values. In particular, each 
hash has bounded moments of all orders (and bounded away from zero), so all conditions of 
the theorem apply.
\begin{align}
\begin{split}
    &\frac{1}{s_n} \sum_{i=1}^n \left( h_r\left(T_i\right) - \mathbb{E}_k\left[h_r\left(T_i\right)\right] \right) \xrightarrow[d]{n \rightarrow \infty} \mathcal{N}(0,1)\\
    \implies &\frac{1}{n} \sum_{i=1}^n \left( h_r\left(T_i\right) - \mathbb{E}_k\left[h_r\left(T_i\right)\right] \right) \xrightarrow[d]{n \rightarrow \infty} \mathcal{N}(0,\frac{S_2 - S_1^2}{n})\\
    \implies &\left(\bar{h} - S_1\right) \xrightarrow[d]{n \rightarrow \infty}  \mathcal{N}(0,\frac{S_2 - S_1^2}{n})\\
    \implies &\bar{h} \xrightarrow[d]{n \rightarrow \infty} \mathcal{N}(S_1,\frac{S_2 - S_1^2}{n})\\
\end{split}
\end{align}

Thus, as $n$ increases, non watermarked text will have $\bar{h}$ get closer to 0.5, while watermarked text will have $\bar{h}$ get closer to $S_1$. In particular, if we fix a maximum false positive ratio $p$ (number of regular text mistaken for watermarked text), the detection strategy is to flag text as watermarked if the probability of $\bar{h}$ in the non-watermarked hypothesis is lower than $p$, or equivalently, $1-\Phi_{0.5,\,1/12n}\left(\bar{h}\right) < p$, with $\Phi_{0.5,\, 1/12n}$ the CDF of a Gaussian centered in 0.5 with variance $1/12n$.

Furthermore, given values of $S_1$ and $S_2$, for large values of $n$, we can compute the expected false negative ratio of our detector. Given the quantile $q$ associated with the false positive ratio $p$ ($\Phi_{0.5,\,1/12n}\left(q\right) = 1-p$), we have $FN = \Phi_{S_1,\,(S_2-S_1^2)/n}(q)$. This gets exponentially lower as the average entropy $S_1$ and $n$ increase.


\section{Societal impact}
\label{app:societal-impact}
Large language models can be misused, which motivates our benchmark for model output watermarking schemes.
We list potential societal impacts of our work below.

\smallskip\noindent\textbf{Designing new watermarks.} Our unified framework for symmetric-key watermarking schemes enables practitioners to build and evaluate custom watermarking schemes using building blocks from different existing work.

\smallskip\noindent\textbf{Deployment readiness.} The results of our benchmark on existing watermarking schemes indicates the need for more work to understand the impact of watermarks on highly structured outputs (e.g., code generation).

\smallskip\noindent\textbf{Regulation.}
Recent legislation from the European Union has placed obligations on providers and users of AI systems to enable the detection and tracing of AI-generated content and to use watermarking schemes at the ``generally acknowledged state of the art''~\cite{eu-ai-act}.
There has yet to be consensus reached on which watermarking scheme is the ``best''. Our benchmark provides a common ground for evaluating watermarking schemes.

Altogether, our work leads to a better understanding of how current model output watermarking schemes perform on real-world use. This can be useful in the development systems that wish to mitigate the risks of misusing large language models.

\end{document}